# Sub-monolayer tracking of molecular states at buried interfaces

Harlan Heilman[1], Em Biegler[1], Sushil Adhikari[1], Victor Murcia[2], Tanner Melody[1], Fred Woodall[1], Lars Thomsen[3], Obaid Alqahtani[2,4], Brian A. Collins[1,2*]
[1]*Department of Physics and Astronomy, Washington State University, Pullman, Washington 99164, USA*
[2]*Department of Materials Science and Engineering, Washington State University, Pullman, Washington 99164, USA*
[3]*Australian Synchrotron, ANSTO, Clayton, VIC, Australia*
[4]*Department of Physics, Prince Sattam bin Abdulaziz University, Alkharj 11942, KSA*

Phenomena responsible for emerging technologies originate from novel molecular states at buried interfaces. Yet quantitatively resolving these features is out of reach in most cases. Resonant soft X-ray scattering (RSoXS) has demonstrated sensitivity to these local states but is limited due to a lack of accurate interpretive models and spatial resolution. Here we demonstrate a machine learning (ML)-optimized optical model that establishes the direct and quantitative interpretation of molecular orientation, chemical identity and packing states with sub-monolayer resolution – reaching beyond the diffraction limit. We discover a gradient evolution of molecular orientation and packing emerging from a pinned state directly at the channel-dielectric interface of an organic transistor, indicative of competing surface versus bulk growth dynamics that contradict models based on substrate patterning. This non-destructive insight into buried interfacial formation and final configuration opens possibilities in controlling interfacial states and phenomena in next-generation technologies.

The molecular orientation[1,2], composition[3] and oxidation state[4,5] at the inaccessible interfaces of buried nanoscale assemblies are decisive quantities that dictate material function critical to advanced technologies. For example, in extreme-ultraviolet photoresists the chemical gradient at the edge of the latent image limits the minimum achievable process size for microprocessors.[6,7] In organic devices such as solar cells, LEDs, transistors, and spin valves the orientation and conformation of the molecular monolayer directly at the interface governs emergent electronic states that control charge generation, recombination, transport, and spin hybridization respectively.[1,2,8–10] In bio- and chemical sensing, molecular states at buried interfaces have resulted in novel ionic transport dynamics.[11–13] Controlling molecular states directly at these interfaces is, therefore, central to advancing each of these technologies.

Unfortunately, the few techniques that resolve buried interfacial structure, such as atom probe tomography or cross-sectional electron microscopy are laborious, highly destructive, and can't resolve details of bonding states or their orientations.[14] Lacking a direct probe, targeted interfacial molecular states (e.g. orientation and packing) in organic devices are thought to be achieved via substrate patterning effects,[15–17] similar to atomic epitaxial growth. Studies have attempted to access the buried molecular state by measuring deposition at coverage checkpoints [18–20] but with disagreement on whether the exposed molecular layers accurately represent the buried interfacial state.

RSoXS, tuned to the energies of electronic transitions into unoccupied molecular orbitals can non-destructively reveal the chemical, electronic, and physical states within these buried structures[7,21,22]. The scattered intensity is determined by two coupled effects, the geometric nanostructure and the optical interaction with the material. For reflectivity analysis, these effects are modeled and fit to scattering data to reveal depth-dependent information.[21,23–27] Unfortunately, the physio-chemical states must be inferred from of optical parameters which increase in number for each energy probed, causing the analysis to become unstable. Therefore, analyses retreat to a single diagnostic resonance which cannot distinguish chemical species nor orientation unless simplifying assumptions are made.[23,28,29] Worse, a buried interfacial layer is typically required to accurately fit the data which is assumed to be comprised of contaminants or oxidation.[23,26,27] It is left undetermined if these layers are real or merely a limitation of the soft X-rays' diffraction limit which sets the smallest resolvable feature by the scatterer's wavelength, in this case, several nanometers.

Recently, near edge X-ray absorption fine structure (NEXAFS) spectroscopy measurements have been used to construct an Edisonian optical model, enabling the orientation state to be inferred, although validity is uncertain and the above limitations remain.[26,27,27] Alternative approaches arrive at the optical model from density functional theory (DFT), yet are limited by fundamental computational inaccuracies.[30–32] Recently, we have demonstrated that ML-refined DFT optical models can quantitatively reproduce angle-resolved (AR) NEXAFS measurements of polycrystalline thin films with different orientation and crystal phases while encoding bond identity and orientation of each absorption resonance.[33]

Here we construct and apply an ML-DFT optical model in RSoXS analysis to directly quantify molecular chemical identity, orientation and density as a function of depth in a molecular thin film/dielectric bilayer used in organic field effect transistors (OFETs). The approach ties spectral resonances together in one physical model of the molecule to increase chemical and orientational sensitivity and locks scattering features between energies and polarization to recover lost diffraction-limited information. This results in sub-monolayer sensitivity of molecular chemical and orientational state. Analysis of the buried interface reveals not contaminants or a substrate patterned crystal phase, but a smooth variation of molecular state with partial coverage of the molecule in a unique orientation and packing state directly at the heterointerface with the dielectric – where the majority of field-effect charge transport occurs in the device.[8] This discovery establishes that competing surface and intermolecular interactions during film growth, rather than substrate templating, set the bulk texture. The new RSoXS analysis, therefore, enables direct characterization of buried structures at the molecular level critical to the physics of crystal growth, kinetic deposition, and solvent interaction models that dictate these interfaces across material systems.

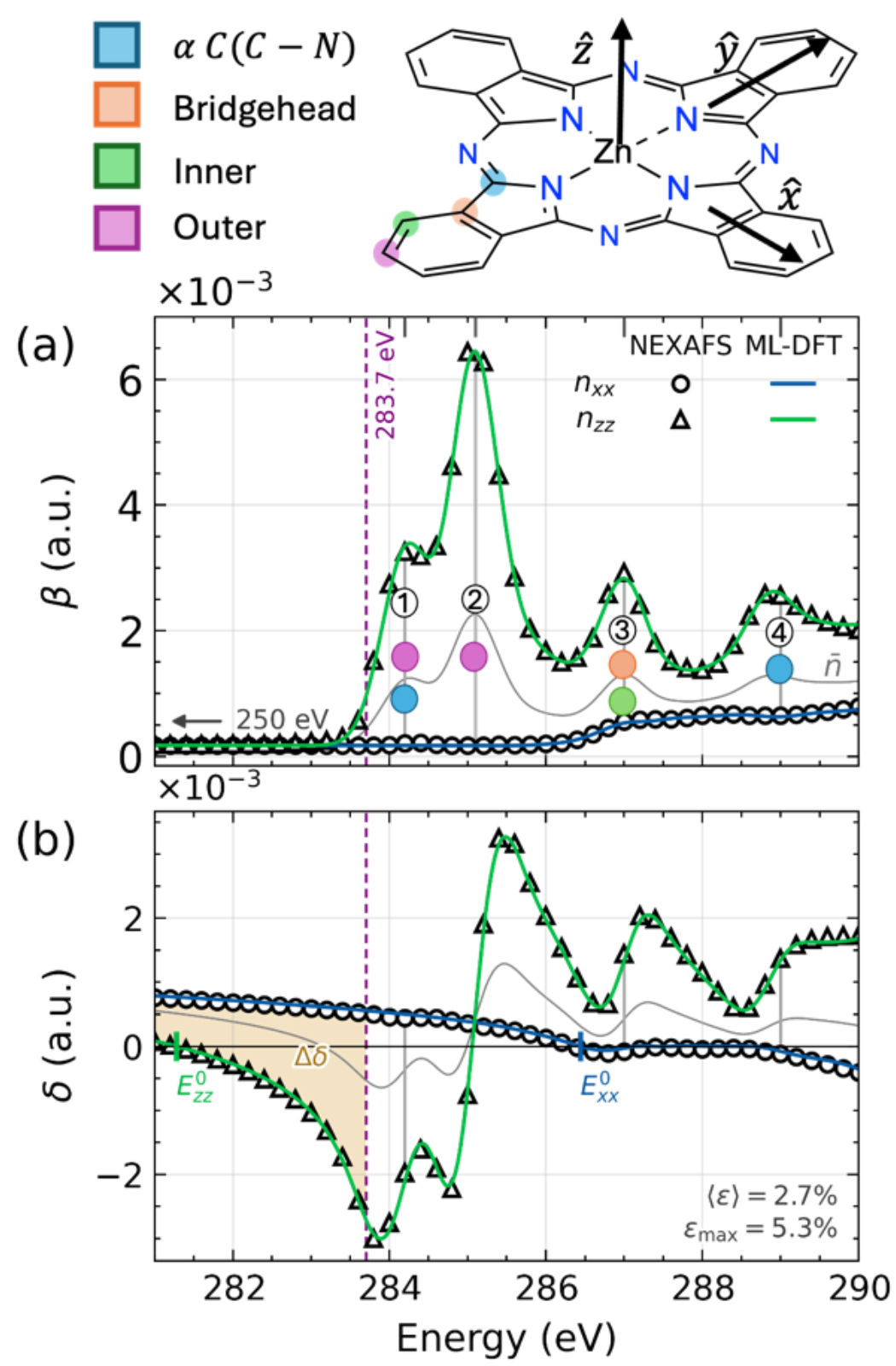

**FIG. 1. An ML-DFT resonant optical model for ZnPc.** ML-DFT (Traces) and angle resolved NEXAFS measurements (points), deviating by $\langle \epsilon \rangle = 2.7\%$ and $\epsilon_{max} = 5.3\%$ at resonance (1). Top, schematic of the molecular frame and the four inequivalent carbon excitation centers, colored as labeled. **(a)** $\beta_{xx}, \beta_{zz}$, and the isotropic average $\bar{\beta}$ (grey), numbered bars are fine structure resonances targeted for RSoXS measurement, with colored pips indicating the excitation centers dominating the transition. **(b)** $\delta$ by Kramers-Kronig transformation[34], with shaded $\Delta\delta$ bounded by the zero crossings $E^0_{xx}$ & $E^0_{zz}$. Vertical purple dash is a diagnostic RSoXS energy.

### Building an ML-refined Optical Tensor

We focus on thin films of Zinc phthalocyanine (ZnPc), a mature molecular platform spanning foundational studies of single-molecule photophysics and quantum optics[34–38], commercial photovoltaics[39,40], thin-film transistors and sensors[41,42], and photomedicines.[43–45] Physical vapor deposition yields polycrystalline films with the molecular orientation influenced by substrate underlayers. CuI and graphene result in face-on orientation (molecules lying flat) with brickstone packing while SiOx patterns edge-on orientation with herringbone packing[15–17]. However, in situ coverage studies have suggested that ZnPc and other metal-Pc films might bear more complex microstructural states that evolve from the substrate.[18–20,46,47] The molecular monolayer at the substrate is particularly important, determining charge transport in in the OFET device.[8]

The monoclinic herringbone crystal structure[17,18] with a strong edge-on texture of ZnPc/SiOx is confirmed by ellipsometry and GIXRD (**SI Sec. 1.1&1.2**). AR-NEXAFS places the film-averaged macrocycle normal ($\hat{z}$ in **Fig. 1** schematic) at a polar tilt of $\gamma = 74°$ from the substrate surface normal (**SI Sec. 1.5**, i.e. edge-on orientation). The molecular index of refraction tensor $n_{ij}(E) = 1 - \delta_{ij}(E) + i\beta_{ij}(E)$ follows by projecting the experimentally measured $\beta(E)$ (from AR-NEXAFS) onto the molecular optical axes (**Fig. 1a**, symbols, **SI Sec. 2**).[48,49] Unfortunately, the measured optical anisotropy is a convolution of the transition dipole moments (TDM) and the film-averaged molecular tilt. Through our ML clustering algorithm we remove this ambiguity to extract the orientation of the principle TDMs calculated from DFT simulated NEXAFS.[33] TDM clusters are then refined to quantitatively reproduce AR-NEXAFS (**SI sec. 3**) compared in **Fig. 1**. The average error between the refined optical tensor in reproducing the AR-NEXAFS data is 2.7% with all experimental features reproduced in **Fig. 1**. Thus, the ML-DFT single molecule tensor (**SI Sec. 4**) fully captures the optical interactions emergent in the film, while assigning chemical character and TDMs to spectral features.

In the model, the first two TDM clustered transitions orient along $\hat{z}$ and carry the dominant oscillator strength, producing a strong pre-edge birefringence ($\Delta\delta$, **Fig. 1b**) that eventually flips at 285 eV. We exploit this in reflectivity measurements via two modes. First, with high-density sampling in energy up to just below the first absorption resonance, we locate the extinction of the critical angle $\theta_C = \sqrt{2\delta}$ to inform the film averaged orientation at low radiation dose. Maximum q-range reflectivity is then collected at two bookending "diagnostic" energies, maximizing sensitivity to thin buried layers. This pre-edge data reports the orientation of a generic conjugated framework rather than chemical identity. Chemical sensing is provided though RSoXS measurements at the fine structure resonances which resolve all four inequivalent carbon environments of the macrocycle (labeled in **Fig. 1** schematic and circles in **Fig.1a**) as determined through the DFT calculation.

### Measuring molecular state vs depth

We analyze the reflectivity data via modeling the film as a three-slab structure (**Fig. 2a)**. Three slabs are required for a good fit and are presumed to capture unique chemical states and species at the surface, bulk and buried interface of the film. We then contrast the ML-DFT optical model with that of the traditional analysis, designated the free model, which fits four uniaxial tensor components per energy per slab (**Fig. 2a**). A secondary analysis is required to determine chemical nature, orientation, and density from these fit components. In contrast, the ML-DFT model directly parameterizes density $\rho$ and molecular tilt angle $\gamma$ (orientation dependence in **Fig. 2b**) with only two parameters per slab compared with 84 in the free model. In this approach, different chemistries, such as oxidation or contamination will be manifest as fit variances at the molecule's key resonant energies (**Fig. 2b & 2e**).

Fit quality between the two models differs by energy regime. Across the pre-edge and resonance series of **Fig. 2d & 2e**, the DFT model tracks the data with similar or better residuals than the free model. The critical angles traced in **Fig. 2d**, vertical black traces, highlight the underlying birefringence that is consistent with the ZnPc molecules oriented edge-on to the substrate. By contrast, the molecule's resonance series (**Fig. 2e**) is sensitive to chemical and electronic orbital identity. Here the ML-DFT model fits better than the free model (average deviation 1.7$\sigma$ vs 2.2$\sigma$, respectively). Since the ML-DFT model restricts only ZnPc to be present in the film, this result indicates a lack of measurable contamination or oxidation in the film to which interfacial layers are customarily assigned.[23,26,27] Instead, these layers depict a unique molecular configurational state well described by the ML-DFT model.

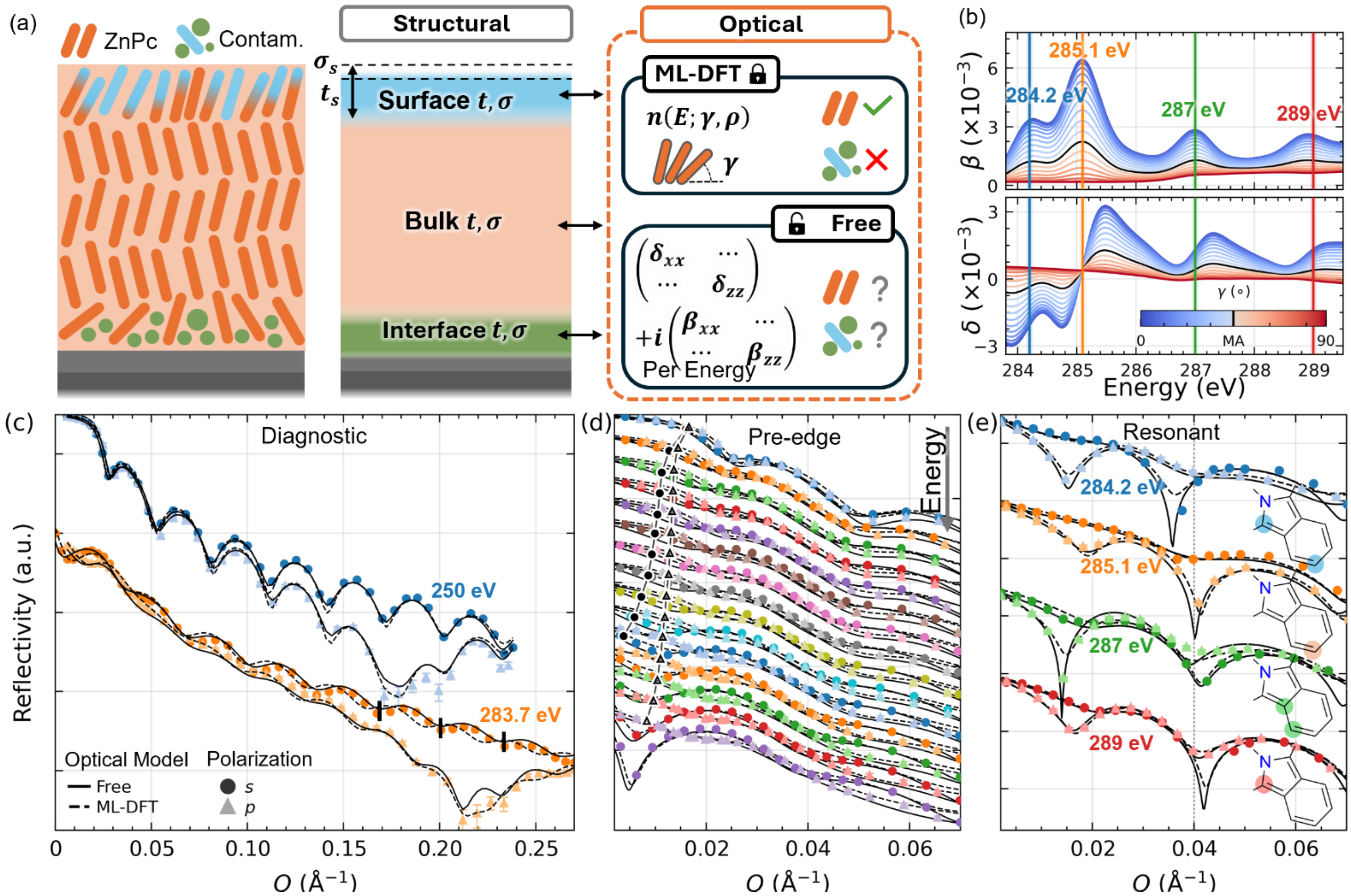

**FIG. 2. Fitting energy and polarization-dependent RSoXS. (a)** Reflectivity model description of the film using three ZnPc slabs on SiO2/Si with thickness $t$ and roughness $\sigma$ each, differing only by the optical parameterization of each slab with the Free model requiring four uniaxial components per energy per slab against the ML-DFT model with just two, molecular tilt $\gamma$ and density $\rho$ per slab. **(b)** Film out of plane tensor components built from **Fig 1** parameterized by tilt angle from face to edge on with the isotropic condition in black. Vertical lines denote the energies in the resonant series. **(c-e)** Reflectivity measurements (showing every fifth datapoint as symbols, **SI sec. 5**),) and model fits (lines) for each energy domain: **(b)** diagnostic energies, **(c)** pre-edge, $\delta$-crossing, **(d)** post-edge, **(e)** resonant. Ticks in **(c)** mark measured fringe minima. Traced in **(d)** plots the critical angle for each s-and-p-pol across the pre-edge energies. Molecule frames inset in **(e)** highlight the carbon excitation centers within one outer ZnPc wing that dominate each TDM cluster.

**Fig. 2c** presents the largest divergences between the models, where the 283.7 eV diagnostic profile is the most challenging to reproduce. At this energy, the ML-DFT model performs worse in the s-polarized profile with a noticeable phase shift (ticks, **Fig. 2c**) between the fringes in the data and model. Conversely, in the p-profile, the free model fits worse and fails to capture the fringe structure around Brewster's angle. Despite this, both models arrive at similar slab structures (**Tab. SI, SII**) that agree with total thicknesses from ellipsometry analysis (**Fig. S1**), yet the ML-DFT model achieves this with >10x sharper parameter uncertainties than the Free model.

At this point, the free model optical parameters must be analyzed further to interpret orientation and density. In contrast, these critical parameters are read directly from the ML-DFT model, quantitatively tracing the microstructure evolution through the film in three distinct densities and orientation states. The parameters of the central bulk layer agree well with tabulated density and orientation of herringbone ZnPc.[15,18] Separately, the surface and interfacial layers exhibit orientation and densities that do not agree with herringbone packing yet both carry the resonant anisotropy that only orientated ZnPc can produce – in particular, the resonant amplitudes and even anisotropy flipping across the edge.

The optical properties of the two models, as shown in **Fig 3a** (expanded in **Fig. S10**), agree closely within the bulk, while those of the surface and buried interface layers are both more extreme and highly uncertain in the free model, resulting in no significant disagreements between the models. Yet the comparison highlights several issues with the free model. For instance, the unphysical dichroism at 287 eV (black arrow pointing off scale in **Fig. 3a**) highlights the fact that there are no constraints against unphysical results. In particular, there is no known way to assure the free model optical parameters obey Kramers-Kronig consistency without excessive energy sampling.[27,50] The second notable issue is interpreting the free model optical parameters for orientational or chemical state. In particular, the low energy trends in **Fig. 3a** are discontinuous

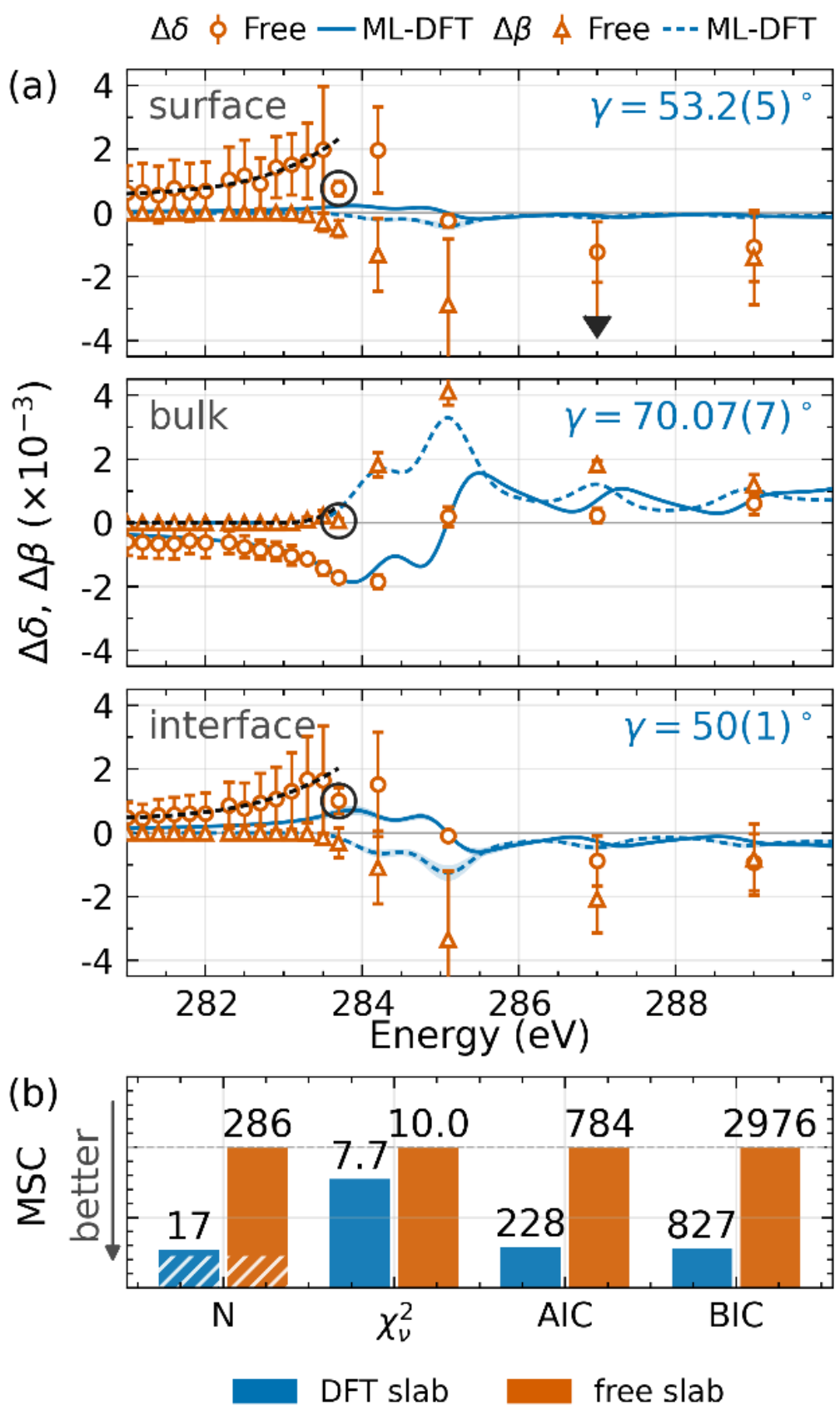

**FIG. 3 Dominance of the ML-DFT model over traditional analyses. (a)** Birefringence $\Delta\delta = \delta_{xx} - \delta_{zz}$ (circles, solid) and dichroism $\Delta\beta = \beta_{xx} - \beta_{zz}$ (triangles, dashed) for the surface, bulk, and interfacial ZnPc layers. Orange points are Free tensor optical parameters fit independently per energy. Blue traces follow from the ML-DFT optical model with blue areas representing uncertainties. **(b)** Model selection criterion (MSC): reduced total parameter counts $N$, reduced $\chi^2$, Akaike information criteria (AIC) and Bayesian information criteria (BIC) for the two models. Lower values indicate the preferred model. Hashed regions in *N* represent constrained experimental offset parameters (**SI Sec. 5.4**)

(circled points), with contradicting conclusions of the molecular orientation depending on which energy is analyzed (**Fig. S10**).

Not only is the ML-DFT model superior, being physically constrained and interpretable, but all statistical metrics demonstrate its advantages (**Fig. 3b**) due especially to the reduced total number of parameters in the model (*N*). From these results, we conclude that the DFT model achieves a drastic statistical improvement in fit quality with 17x fewer parameters (**Si Sec. 5.4**) and 10x smaller parameter uncertainties, while presenting directly interpretable results that are useful to interrogate the molecular state.

### Sub-monolayer interfacial physics

Without needing to account for unique chemical species at the boundaries of the ZnPc film, the discrete slab model is no longer physically motivated. Instead, we develop a single slab gradient model based on the physics of a continuous second order transition from a pinned molecular state due to surface energies that evolve toward an energy-minimized state in the bulk (derivation in **SI Sec. 6**). This results in a description physically matched to an evolving growth front during deposition with each deposited layer taking its orientation from those beneath it. This gradient model resolves orientation as a smooth function of depth, parameterized by the bulk state and two exponentials that transition over characteristic lengths $\tau$ toward the pinned state at each interface. Besides incorporating the physics at play during deposition, this structural model also has fewer parameters than the slab model.

The fit using the gradient model (with ML-DFT optics) is compared to the slab model in **Fig. 4a** for only the 283.7 eV diagnostic profile and the resonant profiles. Notably, the erroneous phase shift of the fringes in the slab model is removed in the gradient model and the Brewster angle is reproduced significantly better than the slab model. These fit improvements along with the reduced number of parameters enable the best goodness of fit, indicating this model recovers the true orientation and density gradient within the sample more accurately than what the slab picture can achieve.

The recovered structure is plotted in **Fig. 4b & 4d** with fit parameters in **Table I**. The bulk parameters are again consistent with edge-on herringbone ZnPc. At the buried interface, the profile descends to a lower density, face-on orientation, while at the free surface the orientation relaxes to a lower angle and a higher density than bulk. This profile is consistent with a second film grown under identical conditions with a different total thickness (**Fig. S10**). The comparison and consistency between independent films suggests these parameters may be intrinsic to the growth conditions of the ZnPc system and dictated directly by the free energy potentials at the vacuum and silica interfaces.

TABLE I. Best fit parameters from the orientation and density gradient model. Total film thickness is 192.7(4) Å with RMS surface roughness of 6.8(4) Å. Errors are reported as $2\sigma$ from the covariance matrix.

| **Region** | $\gamma(\circ)$ | $\rho(g/cm^3)$ | $\tau(\text{Å})$ |
|---|---|---|---|
| **Surface** | 58.7(6) | 2.0(1) | 13(1) |
| **Bulk** | 69.6(2) | 1.64(1) | – |
| **Interface** | 38(1) | 0.87(7) | 14.9(1) |

**Fig. 4d** presents a molecular level, real space reconstruction of the film down to the dielectric interface through data fusion of our RSoXS analysis, XRD lattice spacings (**Fig. S11&12**), and a force-field model of the molecule. Notably, the entire film is only ~15 monolayers thick with both interfacial regions equivalent to a single monolayer in thickness. However, the fact that the gradient interface model fits better than the discrete model strongly indicates that the continuous density and orientation variation within that region represents a real structural variation resolved by this analysis. Such sub-monolayer variation in molecular state in the last few angstroms at the interface, can only manifest with partial coverage of ZnPc molecules that are pinned relatively flat on the substrate. These confined molecular states (~4Å tall, highlighted red in **Fig 4c**), are well below the experimental diffraction limit ($2\pi/q_{max} >$ 23Å). However, scattering from such features still modify the amplitude and phase of the measured Keissig fringes. A Fisher sensitivity analysis (**Fig. S13**) determines that these sub-monolayer features are uniquely captured and constrained by our ML-DFT optical model which locks these effects as a function of energy and polarization. This enables recovery of the molecular state below the diffraction limit.

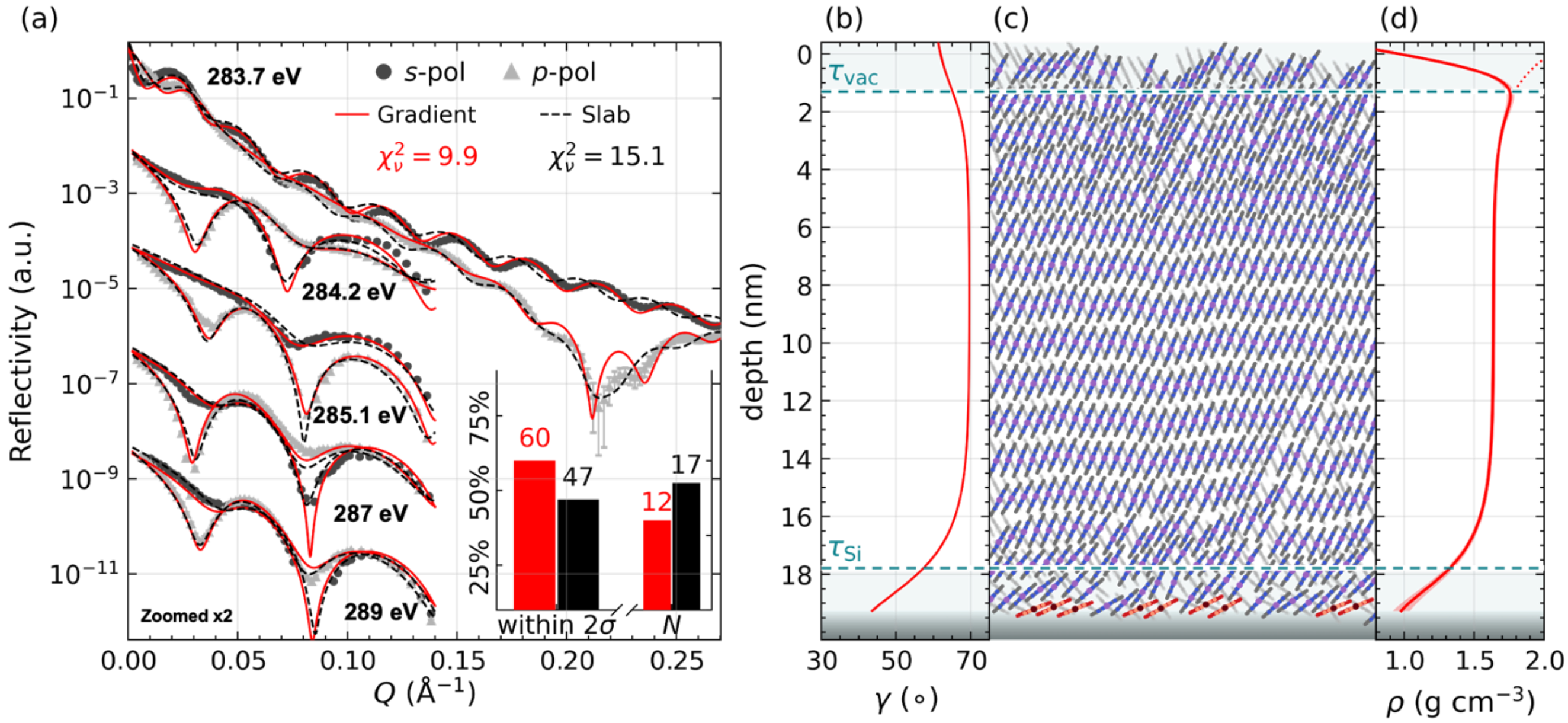

**FIG. 4 Revealing sub-monolayer orientation and density gradients at a buried interface. (a)** s- and p-pol reflectivity fits to 283.7 eV and resonant energies for the slab model (dashed), against the gradient model (red). For clarity, the q-range is stretched by 2x for the resonant profiles. Inset reports model fit quality as $2\sigma$ . **(b-d)** Real space analysis of molecular state versus depth. **(b)** orientation and **(d)** density depth profile of the gradient model with characteristic interfacial widths $\tau_{Si}$ and $\tau_{vac}$ and RMS surface roughness, shaded bands representing 95% confidence ($2\sigma$). **(c)** Visual reconstruction of the molecular state with depth consistent with orientation, density, and diffraction measurements. Grey molecules represent a layer behind the colored molecular layer in the front.

**Surface vs bulk dynamics in film growth**

These results represent evidence against the substrate patterning growth model in favor of an alternative model that incorporates intermolecular interactions at the growth front. One explanation for our results is that the orientation a planar molecule adopts during deposition reflects a competition between the substrate adhesion (favoring face-on alignment) and intermolecular $\pi - \pi$ cohesion (favoring edge-on packing). The gradient then traces the crossover from a low dimensional interfacial coupling to the three-dimensional growth mechanism that sets the bulk texture. The lower density and face-on alignment at the interface indicates that the packing state here is not herringbone but possibly a primordial face-on brickstone packing structure which typically exhibits such properties.[18] Critically, this first monolayer above the dielectric is where the field effect charge transport occurs in the OFET device.[8] Thus, measuring and controlling this unique interfacial state, not the bulk state formed above, is of central importance to device performance. The substrate patterning growth model predicts edge-on herringbone packing at the interface resulting in a significantly different charge carrier density and lateral hopping frequency than the face-on state measured here.

Beyond OFET devices, we expect that evaluating growth dynamics on interfacial states in molecular systems will enable control of emergent phenomena at these interfaces, for example, charge generation and recombination in organic solar cells and OLEDs,[1,2] spin hybridization in organic spin valves,[10] or ion transport in electrochemical devices.[11–13] The same sub-monolayer sensitivity extends directly to self-assembled monolayers, where molecular orientation and coverage is the limiting factor for both efficiency and thermal stability in perovskite solar cells.[9] More broadly the analysis can be applied to any device interface whose function depends on an interfacial structure that a bulk measurement cannot resolve.

The surface densification measured here, along with an orientation approaching the magic angle also indicates a distinct state from the bulk, where the higher density and disorder may be caused by the physics of surface tension during deposition. In this case, the surface layer is not the adventitious hydrocarbon contamination nor oxidation to which film-averaged analyses and other RSoXS studies assign them[27,51] but rather potentially a disordered state before the ZnPc molecules fully crystalize. Moreover, this demonstrates that studies characterizing structure of films deposited at coverage checkpoints[18–20] do not measure the state of a continuously deposited film, since both substrate and free surface energetics pull the state away from the final buried states.

The new RSoXS analysis provides a depth sensitivity to buried molecular state applied in a broad range of physical conditions, allowing quantitative studies of interfacial phenomena like crystal growth, kinetic deposition, and solvent interaction. The same construction extends to molecules with additional orientational and chemical freedom: molecular conformation or orbital hybridization affecting unoccupied electronic orbitals and their TDMs. The analysis is, furthermore, applicable to scattering geometries beyond reflectivity, enabling local analyses of inaccessible buried nanostructures. Connecting scattering features that vary with energy and polarization, our ML-DFT optical model can not only provide these states directly but also resolve structure well beyond the diffraction limit – removing a traditional disadvantage for soft X-rays.

In summary, we introduce a resonant X-ray scattering analysis that directly interprets the quantitative physiochemical molecular state in a buried nanostructure based on an ML-refined DFT optical model. The analysis dramatically increases the parameter precision and extends sensitivity below the diffraction limit, so that molecular state at an interface can be directly characterized. Applying this analysis to ZnPc/SiOx transistor bilayers, we find that hypotheses of chemical contamination at the molecular film surface and buried interface are rejected in favor of the molecules exhibiting a novel packing and orientational state. Introducing a structural model based on the dynamics of film growth, we reveal evidence to reject hypotheses of substrate patterning of the bulk packing state replaced with a competition between 2D versus 3D film growth dynamics that determines the interfacial evolution of the molecular states. This statistical analysis of the final buried molecular state contrasts with checkpoint studies that track a growth front that is only in a temporary state. We envision this new non-destructive view inside buried interfaces and structures will lead to exquisite control of novel functionalities.

## Online content

Any methods, additional references, summaries, source data, extended data, supplementary information, acknowledgements; details of author contributions and competing interests; and statements of data and code availability are available online

1. Tumbleston, J. R. *et al.* The influence of molecular orientation on organic bulk heterojunction solar cells. *Nature Photon* **8**, 385–391 (2014).
2. Yokoyama, D. Molecular orientation in small-molecule organic light-emitting diodes. *J. Mater. Chem.* **21**, 19187–19202 (2011).
3. Collins, B. A. *et al.* Absolute Measurement of Domain Composition and Nanoscale Size Distribution Explains Performance in PTB7:PC71BM Solar Cells. *Advanced Energy Materials* **3**, 65–74 (2013).
4. Cabana, J., Kwon, B. J. & Hu, L. Mechanisms of Degradation and Strategies for the Stabilization of Cathode–Electrolyte Interfaces in Li-Ion Batteries. *Acc. Chem. Res.* **51**, 299–308 (2018).
5. Wang, L. *et al.* Insights into the Dynamic Surface Reconstruction of Electrocatalysts in Oxygen Evolution Reaction. *Renewables* **2**, 272–296 (2024).
6. Bisschop, P. D. Stochastic effects in EUV lithography: random, local CD variability, and printing failures. *JM3* **16**, 041013 (2017).
7. Andrle, K., Zhang, Q., Heilman, H., Ferron, T. & Wang, C. Nanostructure characterization via RSoXS: inverse scattering with FEM-based Maxwell solvers. in *Metrology, Inspection, and Process Control XL* vol. 13981 139810D (SPIE, 2026).
8. Pei, M. *et al.* Semiconductor/dielectric interface in organic field-effect transistors: charge transport, interfacial effects, and perspectives with 2D molecular crystals. *Advances in Physics: X* **5**, 1747945 (2020).
9. Jiang, W. *et al.* Toughened self-assembled monolayers for durable perovskite solar cells. *Nature* **646**, 95–101 (2025).
10. Droghetti, A. *et al.* Dynamic spin filtering at the Co/Alq3 interface mediated by weakly coupled second layer molecules. *Nat Commun* **7**, 12668 (2016).
11. Khan, T., McAfee, T., Ferron, T. J., Alotaibi, A. & Collins, B. A. Local Chemical Enhancement and Gating of Organic Coordinated Ionic-Electronic Transport (Adv. Mater. 5/2025). *Advanced Materials* **37**, 2570041 (2025).
12. Trigg, E. B. *et al.* Self-assembled highly ordered acid layers in precisely sulfonated polyethylene produce efficient proton transport. *Nature Mater* **17**, 725–731 (2018).
13. Dong, B., Gwee, L., Salas-de la Cruz, D., Winey, K. I. & Elabd, Y. A. Super Proton Conductive High-Purity Nafion Nanofibers. *Nano Lett.* **10**, 3785–3790 (2010).
14. Cojocaru-Mirédin, O. *et al.* Atom Probe Tomography: a Local Probe for Chemical Bonds in Solids. *Advanced Materials* **36**, 2403046 (2024).
15. Kato, M. *et al.* Morphological and optical properties of α- and β-phase zinc (Ⅱ) phthalocyanine thin films for application to organic photovoltaic cells. *J. Chem. Phys.* **153**, 144704 (2020).
16. Ngongang Ndjawa, G. O. *et al.* Impact of Molecular Orientation and Spontaneous Interfacial Mixing on the Performance of Organic Solar Cells. *Chem. Mater.* **27**, 5597–5604 (2015).
17. Schünemann, C. *et al.* Zinc phthalocyanine — Influence of substrate temperature, film thickness, and kind of substrate on the morphology. *Thin Solid Films* **519**, 3939–3945 (2011).
18. Derrien, T. L. *et al.* Interfacial rearrangements and strain evolution in the thin film growth of ZnPc on glass. *Phys. Rev. Mater.* **6**, 033401 (2022).
19. Yamamoto, I., Imamura, M., Azuma, J. & Takahashi, K. Variation in zinc phthalocyanine molecular orientation on SrTiO3(100) surface. *Surface Science* **726**, 122162 (2022).
20. de Oteyza, D. G., Barrena, E., Sellner, S., Ossó, J. O. & Dosch, H. Structural Rearrangements During the Initial Growth Stages of Organic Thin Films of F16CuPc on SiO2. *J. Phys. Chem. B* **110**, 16618–16623 (2006).
21. Grabner, D., Heilman, H., Patterson, A., Melody, T. M. & Collins, B. A. Resonant Soft X-Ray Scattering in Polymer Materials. *Annual Review of Materials Research* **55**, (2025).
22. Andrle, K. *et al.* Using a FEM-based Maxwell solver and CD-RSoXS measurement for the analysis of photoresist pattern. in *Advances in Patterning Materials and Processes XLII* vol. 13428 134281G (SPIE, 2025).
23. Ferron, T. J. *et al.* Characterization of the Interfacial Orientation and Molecular Conformation in a Glass-Forming Organic Semiconductor. *ACS Appl. Mater. Interfaces* **14**, 3455–3466 (2022).
24. Wang, C. *et al.* Resonant soft x-ray reflectivity of organic thin films. *Journal of Vacuum Science & Technology A: Vacuum, Surfaces, and Films* **25**, 575–586 (2007).
25. Yan, H., Wang, C., McCarn, A. R. & Ade, H. Accurate and Facile Determination of the Index of Refraction of Organic Thin Films Near the Carbon 1 s Absorption Edge. *Phys. Rev. Lett.* **110**, 177401 (2013).
26. Mezger, M. *et al.* Molecular orientation in soft matter thin films studied by resonant soft x-ray reflectivity. *Phys. Rev. B* **83**, 155406 (2011).
27. Ferron, T. J., Fiori, M. E., Ediger, M. D., DeLongchamp, D. M. & Sunday, D. F. Composition Dictates Molecular Orientation at the Heterointerfaces of Vapor-Deposited Glasses. *JACS Au* **3**, 1931–1938 (2023).
28. Li, H. *et al.* Pattern-enhanced Resonant Soft X-ray Scattering for Operando monitoring of electrochemical solid-liquid interfaces. *Nat Commun* https://doi.org/10.1038/s41467-026-69852-9 (2026) doi:10.1038/s41467-026-69852-9.
29. Nguyen, P. H. *et al.* Resonant Soft X-ray Scattering Reveals the Distribution of Dopants in Semicrystalline Conjugated Polymers. *J. Phys. Chem. B* **128**, 12597–12611 (2024).
30. Pasquali, L. *et al.* Structural and electronic properties of anisotropic ultrathin organic films from dichroic resonant soft x-ray reflectivity. *Phys. Rev. B* **89**, 045401 (2014).
31. Capelli, R. *et al.* Quantitative resonant soft x-ray reflectivity from an organic semiconductor single crystal. *J. Chem. Phys.* **150**, 094707 (2019).
32. Pasquali, L. *et al.* Analysis of Resonant Soft X-ray Reflectivity of Anisotropic Layered Materials. *Surfaces* **4**, 18–30 (2021).
33. Murcia, V., Alqahtani, O., Heilman, H. & Collins, B. A. Quantitative and Bond-Traceable Resonant X-Ray Optical Tensors of Organic Molecules. *Phys. Rev. Lett.* **137**, 158001 (2026).

34. Yang, B. *et al.* Sub-nanometre resolution in single-molecule photoluminescence imaging. *Nat. Photonics* **14**, 693–699 (2020).
35. Zhang, Y. *et al.* Visualizing coherent intermolecular dipole–dipole coupling in real space. *Nature* **531**, 623–627 (2016).
36. Luo, Y. *et al.* Electrically Driven Single-Photon Superradiance from Molecular Chains in a Plasmonic Nanocavity. *Phys. Rev. Lett.* **122**, 233901 (2019).
37. Cao, S. *et al.* Energy funnelling within multichromophore architectures monitored with subnanometre resolution. *Nat. Chem.* **13**, 766–770 (2021).
38. Doppagne, B. *et al.* Electrofluorochromism at the single-molecule level. *Science* **361**, 251–255 (2018).
39. Klenkler, R. A., Yuen, A. P. & Bamsey, N. M. Photovoltaic device. (2015).
40. Pfeiffer, M. *et al.* Organic photoactive device. (2013).
41. Boileau, N. T., Cranston, R., Mirka, B., Melville, O. A. & Lessard, B. H. Metal phthalocyanine organic thin-film transistors: changes in electrical performance and stability in response to temperature and environment. *RSC Adv.* **9**, 21478–21485 (2019).
42. Comeau, Z. J. *et al.* Surface engineering of zinc phthalocyanine organic thin-film transistors results in part-per-billion sensitivity towards cannabinoid vapor. *Commun Chem* **5**, 178 (2022).
43. Chornovolenko, K. & Koczorowski, T. Phthalocyanines Conjugated with Small Biologically Active Compounds for the Advanced Photodynamic Therapy: A Review. *Molecules* **30**, 3297 (2025).
44. Rak, J., Kabesova, M., Benes, J., Pouckova, P. & Vetvicka, D. Advances in Liposome-Encapsulated Phthalocyanines for Photodynamic Therapy. *Life* **13**, 305 (2023).
45. Luo, T. *et al.* Nanoscale Metal–Organic Framework Confines Zinc-Phthalocyanine Photosensitizers for Enhanced Photodynamic Therapy. *Journal American Chemical Society* **143**, 13519–13524 (2021).
46. Yang, J., Yim, S. & Jones, T. S. Molecular-Orientation-Induced Rapid Roughening and Morphology Transition in Organic Semiconductor Thin-Film Growth. *Sci Rep* **5**, 9441 (2015).
47. Ossó, J. O. *et al.* Structure, morphology, and optical properties of thin films of F16CuPc grown on silicon dioxide. *Organic Electronics* **5**, 135–140 (2004).
48. Stöhr, J. *NEXAFS Spectroscopy*. (Springer Berlin, Heidelberg, 1992).
49. Collins, B. A. *et al.* Polarized X-ray scattering reveals non-crystalline orientational ordering in organic films. *Nature Mater* **11**, 536–543 (2012).
50. Stone, K. H., Valvidares, S. M. & Kortright, J. B. Kramers-Kronig constrained modeling of soft x-ray reflectivity spectra: Obtaining depth resolution of electronic and chemical structure. *Phys. Rev. B* **86**, 024102 (2012).
51. Fostiropoulos, K., Mete, T., Li, Z. & Lauermann, I. Temperature-induced molecule assembly effects on the near-edge X-ray absorption fine-structure spectra of Zn-phthalocyanine layers on Si substrates. *J Synchrotron Rad* **26**, 1260–1265 (2019).
52. Kozlik, M., Paulke, S., Gruenewald, M., Forker, R. & Fritz, T. Optical Constants of α- and β-Zinc(II)-Phthalocyanine Films. *Dataset Papers in Science* **2013**, 926470 (2013).
53. Ilavsky, J. *Nika*: software for two-dimensional data reduction. *J Appl Crystallogr* **45**, 324–328 (2012).
54. Hoshino, A., Takenaka, Y. & Miyaji, H. Redetermination of the crystal structure of α-copper phthalocyanine grown on KCl. *Acta Cryst B* **59**, 393–403 (2003).
55. Nečas, D. & Klapetek, P. Gwyddion: an open-source software for SPM data analysis. *Open Physics* **10**, 181–188 (2012).
56. Heilman, H., Fred, W., Alqahtani, O., Lars, T. & Collins, B. A. X-ray atlas NEXAFS dataset: Zn-phthalocyanine, C K, total electron yield, SXR, the australian synchrotron. X-ray Atlas https://doi.org/10.5281/zenodo.22742990 (2026).
57. Henke, B. L., Gullikson, E. M. & Davis, J. C. X-Ray Interactions: Photoabsorption, Scattering, Transmission, and Reflection at $E$ = 50-30,000 eV, $Z$ = 1-92. *Atomic Data and Nuclear Data Tables* **54**, 181–342 (1993).
58. Mai, T. Data for: Density Function Theory Calculation, and Phthalonitrile Process for a Synthesis of Single Crystal Zinc Phthalocyanine. Mendeley https://doi.org/10.17632/TX5H3SRHZT.1 (2020).
59. Victor Murcia & Harlan D Heilman. wsu-carbon-lab/dft-learn. Zenodo https://doi.org/10.5281/ZENODO.14758696 (2025).
60. als-rsoxs/pxr-reduce. ALS RSOXS https://doi.org/https://doi.org/10.5281/zenodo.22759048 (2026).
61. als-rsoxs/refloxide. ALS RSOXS https://doi.org/https://doi.org/10.5281/zenodo.22759138 (2026).
62. Névot, L. & Croce, P. Caractérisation des surfaces par réflexion rasante de rayons X. Application à l'étude du polissage de quelques verres silicates. *Rev. Phys. Appl. (Paris)* **15**, 761–779 (1980).
63. Nelson, A. R. J. & Prescott, S. W. refnx: neutron and X-ray reflectometry analysis in Python. *J Appl Cryst* **52**, 193–200 (2019).
64. Thompson, A. P. *et al.* LAMMPS - a flexible simulation tool for particle-based materials modeling at the atomic, meso, and continuum scales. *Computer Physics Communications* **271**, 108171 (2022).
65. Heilman, H. refl-analysis. https://github.com/WSU-Carbon-Lab/refl-analysis (2026).
66. Schünemann, C. *et al.* Evaluation and Control of the Orientation of Small Molecules for Strongly Absorbing Organic Thin Films. *J. Phys. Chem. C* **117**, 11600–11609 (2013).
67. Chikamatsu, T. *et al.* Identifying Molecular Orientation in a Bulk Heterojunction Film by Infrared Reflection Absorption Spectroscopy. *ACS Omega* **3**, 5678–5684 (2018).
68. CXRO X-Ray Interactions With Matter. https://henke.lbl.gov/optical_constants/.
69. Watts, B. Calculation of the Kramers-Kronig transform of X-ray spectra by a piecewise Laurent polynomial method. *Opt. Express, OE* **22**, 23628–23639 (2014).
70. Esashi, Y. *et al.* Influence of surface and interface roughness on X-ray and extreme ultraviolet reflectance: A comparative numerical study. *OSA Continuum, OSAC* **4**, 1497–1518 (2021).
71. Gennes, P.-G. de & Prost, J. *The Physics of Liquid Crystals*. (Clarendon press, Oxford, 1993).
72. Ginzburg, V. L. & Landau, L. D. On the theory of superconductivity. in *On superconductivity and superfluidity: a scientific autobiography* 113–137 (Springer Berlin Heidelberg, Berlin, Heidelberg, 2009). doi:10.1007/978-3-540-68008-6_4.

## Methods

**Film growth:** Small molecule ZnPc (C32H16N8Zn sourced from Sigma Aldrich) with purity ≥ 99.5% was further purified in the physical vapor deposition (PVD) chamber (base pressure $< 1 \times 10^{-7}$ Torr) by gradually heating, approaching ZnPc sublimation temperature over approximately an hour until no further off gassing of contaminants were detected by the quartz crystal monitor (QCM). Thin films (~20 nm measured by QCM) of ZnPc were deposited at $0.2$ Å$/s$ on high conductivity polished Si wafers (University Wafers). Wafers were cleaved 1 cm x 3 cm and cleaned via sonication in deionized water, acetone and isopropyl alcohols before a 30 min UV ozone bake. ZnPc was either deposited directly onto these wafers to create the herringbone crystal structure, with the substrate temperature set to the ambient room temperature (24 C). Films for RSoXS were stored under an inert N2 atmosphere and transported to the ALS under vacuum. Auxiliary pre characterization measurements were performed on separate films from the same batch to avoid surface contamination and oxidation prior to RSoXS.

**VASE:** Variable angle spectroscopic ellipsometry (VASE) (**Fig S1**) was performed using a J.A. Woollam alpha-SE to pre characterize layer thickness and molecular orientation. Data was collected at three incidences ($65°, 70°, \& 75°$) targeting a sample from the same PVD batch and measured using vendor supplied CompleteEase software. Optical constants where modeled using a B-spline producing similar spectra to literature.[52] The results and analysis of VASE data are discussed in **SI Sec. 1.1.**

**GIXRD:** Grazing incidence wide angle x-ray diffraction (GIXRD) measurements (**Fig S2**) were performed at Washington State University on a Xenocs Xeuss 3.0 with a focused Cu X-ray source ($\lambda = 1.5418$ Å), under vacuum at 72 mm sample-detector distance with incident angle set to $0.2°$. The sample-to-detector distance and beam center were calibrated with a silver-behenate standard. 1D reduction of GIWAXS images was done using NIKA software [53] and peak positions were obtained by fitting Gaussian line shapes to the reduced profiles. The lamellar reflection lies along the $q_z$ and the $\pi - \pi$ reflection is in plane corresponding with the edge on texture of the $\alpha$ herringbone polymorph reported for vacuum deposited ZnPc and CuPc.[54] The results and analysis of GIXRD data are discussed in **SI Sec. 1.2**.

**AFM:** Atomic force microscopy (AFM) height maps were acquired on a Bruker Dimension Icon in tapping mode with a TESPA-V2 SAMPLE probe. Scans where processed using Gwyddion[55] to extract root-mean-square (RMS) roughness values. Film morphologies and RMS roughness values were used to pre-characterize film surface roughness and morphology RSoXS measurements. The surface topology is included in **Fig S4** and the discussion in **SI Sec. 1.3** and **1.4**.

**NEXAFS spectroscopy:** NEXAFS spectra were collected at the Soft X-ray Spectroscopy (SXR) beamline of the Australian Synchrotron, part of ANSTO in total electron yield mode on Si wafers.[56] Samples where grown under the same conditions Spectra were collected at five sample alignments with respect to the incident electric field polarization vector with respect to the surface normal ($20°, 40°, 55°, 70°, \& 90°$). NEXAFS was collected over the energy range of 270-350 eV and normalized to the energy dependent intensity fluctuation and optical loss via a direct photodiode measurement, and Au mesh drain current. Mass absorption was computed via normalization in the pre and post edge regions to the bare atom mass absorption of ZnPc as extracted from the CXRO Henke atomic scattering factors.[57]

Molecular tilt angle was extracted via the DFT cluster and refinement process. As the final step of the refinement process, the molecular polar tilt angle $\gamma$ is fit using the TDM cluster orientations prescribed by DFT providing a measure of the molecular orientation (**Fig S5**). This fit produces a reduces $\chi^2$ goodness of fit statistic of .213 indicating very good agreement between the model and experimental data. The molecular tilt angle of 74° agrees with all other characterizations of the ZnPc edge on herringbone polymorph described in this study and is taken as the approximate orientation within the films measured under RSoXS (full analysis in **SI Sec. 3**).

**Density Functional Theory:** A time independent transition potential calculation using the StoBe library was performed on a single molecule of ZnPc. A HCH approach sets the manifold of transition energies, along with each TDM and oscillator strength, while the $\Delta KS$ method is used to fix the absolute energy associated with each transition. For details see our previous work.[33] The molecular geometry was extracted from a crystallographic structure file (.cif)[58] as it produced results with a higher initial agreement to experiment compared to thermodynamically relaxed structures, or those relaxed from molecular force fields. The final calculation results are provided in **SI Sec. 1.6**.

**ML-clustering and refinement:** The ML optimized optical tensor is constructed though hierarchical agglomerative clustering to extract principle TDM clusters. Exact implementation details may be found on GitHub or through our previous work.[33,59] TDM clusters are sequentially refined to quantitative agreement with experimental AR-NEXAFS by fitting oscillator strength, excitation energy, and TDM orientation. Optimal hyperparameters, oscillator strength threshold and overlap percentage were investigated by evaluating this best fit metric on an even grid $OS \in [0, 10]$ & $OVP \in [25,75]$ with details in **SI Sec 3.1** and highlighted in **Fig S7**. From the family of hyperparameters that perform well on this metric, the refinement performance is additionally weighted by the adherence to the original DFT spectra and TDM cluster families such that the refinement step does not significantly alter the DFT calculation. Here we chose the best clustering parameters of $OS = 0.25$ and $OVP = 52.5$ with the full evaluation matrix in **Fig. S7**. In general, we find good clustering and convergence for all $OVP \in [55,50]$, with significantly worse performance outside of these bounds. Moreover, we find that $OS \in [.1,8]$ performed similarly, yet parameter agreement with originating DFT degrades rapid for larger $OS$, while the refinement never converse for $OS < .1$.

**RSoXS collection:** Polarized resonant soft x-ray reflectivity was measured at beamline 11.0.1.2 of the Advanced Light Source, Lawrence Berkeley National Laboratory, following the procedure established by Ferron et al.[27] Samples are transferred from inert storage to the end station where specular reflectance is captured on the CCD area detector with thermals managed by thermos-electric cooling. The sample surface is aligned to the goniometer axis of rotation by an iterative process that refines the sample rotation and height. Incident flux was attenuated using the scatter slits together with the higher order suppressor to avoid detector damage. Because the total reflectance varies by more than the detectors dynamic range, reflectivity is captured in overlapping $\boldsymbol{q}$ subsets with suppressor settings and exposure time raised at each boundary once reflectivity reaches the noise floor.

Data is collected first in s polarization and then in p polarization beginning at 250 eV below the carbon k edge and stepping upward though the resonance, for 21 photon energies collected in batches of five energies with the axis of rotation alignment was reevaluated between energy batches. Upon polarization change, optics were allowed to re equilibrate to the new heat load for 30 min before alignment was recalibrated. The complete dataset was acquired within a single 12 h shift. Photon energy was calibrated against the transmission NEXAFS of a polystyrene standard at the beginning and at the end of the shift with no discernable shift, bounding the energy drift over the experiment to the width of the primary polystyrene feature.

**RSoXS reduction:** Reflectance was reduced from the two-dimensional detector images using the RSoXS-reduce package.[60] The specular spot was located on each frame and its intensity integrated over a fixed region of interest. An equal region displaced along the same detector row provided the background, so that subtraction removes the systematic row-wise readout offset in addition to dark current and stray light. Reflectance was then formed by dividing the background-corrected specular intensity by the incident $I_0$ recorded at the start of each reflectivity profile through the averaging of five direct beam measurements.

At each energy, full multi-decade reflectivity profiles are constructed by properly scaling and joining the overlapping $\boldsymbol{q}$-subsets measured at different attenuation and exposure time. An effective $I_0$ is computed from the four overlapping points, and the subsets were joined to form a singular continuous reflectivity profile for each measured energy. Reported uncertainties combine per-pixel effective $I_0$ with a $\boldsymbol{q}$-dependent envelope accounting for the increasing ambiguity between specular and diffuse reflectivity at higher $\boldsymbol{q}$ using the following equation:

$$\sigma(E, \boldsymbol{q}) = aR_{exp}(E, \boldsymbol{q}) + b\boldsymbol{q} \quad (5)$$

With $a = .08$ and $b = .5 \times 10^{-6}$, which is a feasible approximation for experimental error under these considerations (discussion in **SI Sec. 5.3**).

**RSoXS structural and optical modeling:** Reflectivity is computed from the 4x4 transfer matrix using the refloxide package.[61] The film was modeled as three ZnPc slabs, denoted surface, bulk and interface, on a native $SiO_2$ layer and a semi-infinite Si substrate. Thickness and roughness of each ZnPc slab together with the oxide thickness and roughness give eight structural parameters. Reflection and transmission coefficients were computed at each photon energy and scattering vector with the 4x4 transfer matrix formalism matching boundary conditions at the five interfaces, with interfacial toughness entering in the Nevot-Croce sense as an exponential attenuation representing off-specular diffuse scattering from a faceted interface.[62] Vacuum, silicon, silicon dioxide, and silicon where assigned to tabulated bare atom optical constants,[57] with an energy-independent density allowed to vary within (2, 2.8) g $cm^{-3}$ to accommodate stochiometric deviation from the tabulated values. The forward model was evaluated using refloxide, and the objective and parameter machinery where provided by refnx.[63]

Two optical parameterizations were compared on this identical structure. The free tensor parameterization assigns each ZnPc slab an independent birefringence, dichroism, isotropic $\beta$ and isotropic $\delta$ at every measured energy, constraining to ranges dictated by the starting optical model, amounting to 252 optical tensor components. The DFT parameterization instead describes each slab energy-independently by a molecular orientation and a density, with one energy offset term common to all layers absorbing the calibrated difference between the NEXAFS and RSoXS instruments, amounting to seven parameters for all three slabs. A detailed parameter accounting and comparison with prior art is discussed in **SI Sec. 5.4**.

Experimental corrections were applied per energy and per polarization as a normalization scale factor and an angle-of-incidence offset, following Ferron et al.,[27] adding four terms per energy at 84 in total. The refinement therefore varies 101 parameters for the DFT model and 350 for the free tensor model. Bound on the correction terms, and the correlations they induce are described in **SI Sec. 5.4**.

All energies and polarizations are co-refined against a single global objective, so that the structural parameters are shared across the dataset and only the optical and correction terms carry energy indices. The log-likelihood assumed Gaussian errors at the uncertainties propagated though reduction, and all varying parameters were assigned uniform priors over their stated bounds. The maximum likelihood parameter set was located by differential evolution over the full bounded space, in parallel with deferred population updating and a local polish step using the L-BFGS-B algorithm, avoiding dependency on initial conditions. Reported parameter uncertainties are then taken from the inversion of the Hessian evaluated in the final polish step. Model comparison uses reduced chi square together with the Akaike and Bayesian information criteria.

**Unit cell construction:** Unit cells used for molecular dynamics simulations are constructed though fusion of GIXRD $d$-spacings with literature unit cells. The rigid ZnPc molecule originates from the same $\beta$-ZnPc (**Fig S13a**) crystallographic results used for DFT simulations.[58] Lennard-Jones parameters are GAFF2, with UFF for the Zn coordination center. Charges are fitted using the B3LYP/def2-SVP electrostatic potentials and used to set the electrostatic repulsion between molecules.

The $\beta$-ZnPc stable polymorph with a known structure serves only as the validation target. A suitable unit cell was constructed from the chemically similar $\alpha$-CuPc by placing rigid ZnPc molecules within the CuPc unit cell. Lattice parameters **a** and **c** are set by the lamellar spacing (002), with **b** set by the $\pi - \pi$ packing. The molecular tilt angle is scanned and compared to the GIXRD patterns though a fiber-texture forward model, with other unit cell angles set by minimizing the lattice energy (UFF+ESP). The resulting lattice parameters are tabulated in **Table SIII** with the resulting crystallographic (.cif) structure available and visualized in **Fig S12b**. The final unit cell is then thermally relaxed by rigid-body NPT at 300K, with independent barostats on each axis such that the force fields set the lattice and herringbone arrangements. Over this range the minimum enthalpy cell produces a tilt of $72(1)°$ and a density of $1.657$ g/$cm^3$ against $72.9(4)°$ and $1.64(3)$ g/$cm^3$ from RSoXS and reproduces the lamellar and (002) spacings within 1.5%. The final relaxed unit cell is visualized in **Fig S12c** used in the remaining MD simulations.

**Real space molecular visualization:** Molecular packing in **Fig. 4c** is a molecular dynamics (MD) reconstruction of the film. It is built from the fitted depth profiles of **Fig 4a,b** and relaxed in LAMMPS.[64] ZnPc is treated as a rigid body in the molecular frame used for the DFT calculations. Lennard-Jones parameters are taken from GAFF2, with UFF for Zn, and atomic charges are fitted to the B3LYP/def2-SVP electrostatic potential. This force field reproduces the $\beta$-ZnPc unit cell at 100K to within 1.5% on each axis (SI).

The film is constructed for pure visualization by data fusion with the fitted depth profiles, GIXRD, and the fitted electrostatic potentials. The molecular frame is the same used for the DFT calculations (C32H16N8Zn, $M = 577.9$ g/mol). It's van der Waals envelope gives an in-plane diameter of 14.5 Å and a face-to-face thickness of 3.6 Å, and the $\pi$-stacking separation is taken as 3.4 Å. At each depth $z$ the molecular tilt is $\gamma(z)$ from the orientation profile, and the area per molecule in the drawing plane is $M/(\rho(z)N_A t)$, with $\rho(z)$ the fitted density profile and $t = 14.4$ Å the extrusion thickness of the slab that the projection represents. The packing density of the molecules in the illustration is therefore self-consistent with the density and orientation gradients from **Fig 4b&d**. The projected cell is orthogonal, carrying a lamellar period along $z$ and a column period in plane whose ratio is fixed by the molecular diameter, the $\pi$ stacking separation and $\sin^2(\gamma(z))$. The lamellae period is modified by the interdigitation between neighboring lamellae, set by a dimensionless factor $f = 0.96$, so adjacent layers can overlap by 4% of the molecular diameter. This together with the herringbone splay and positional and orientational disorder are set only for visual continuity and are neither empirical nor derived factors.

The depth dependent lamellae and $\pi$ d-spacings are compared with diffraction orders derived from experimental GIWAXS measurements in **Fig S11**. Because both the cell area and its aspect ratio vary with depth, each z monolayer contributed its own pair of spacings rather than a single value, so the predicted diffraction positions are a taken as the maxima of these depth distributions, weighted by the local density and the Debye-Waller factor built from the drawn positional disorder. The lamellar and $\pi$ spacing periods agree with the measured peak centers to around 2%, treated as a secondary consistency check of the molecular visualization.

**Data Availability:** NEXAFS data that support the findings of this study have been deposited in X-ray Atlas (https://xrayatlas.wsu.edu) with the accession code 193bc0c5 and are archived at Zenodo (https://doi.org/10.5281/zenodo.22742990).[56] RSoXS reflectivity curves, GIWAXS, non-resonant XRR, VASE, and AFM topography together with the fitted model parameters that support the findings of this study are available from the corresponding author upon reasonable request.

**Code availability:** The source code used for reflectivity analysis may be found on GitHub (https://github.com/wsu-carbon-lab/refl-analysis).[65] The code for the RSoXS 1D to 2D data reduction can be found on GitHub (https://github.com/als-rsoxs/pxr-reduce).[60] The source code used to compute reflectivity and fit RSoXS data may be found on GitHub (https://github.com/als-rsoxs/refloxide).[61] The source code used for the DFT clustering and optical model generation may be found on GitHub (https://github.com/wsu-carbon-lab/dft-learn).[59]

**Acknowledgements:** This research used resources of the Advanced Light Source, which is a DOE Office of Science User Facility under contract no. DE-AC02-05CH11231. Part of this work was carried out on the Soft X-ray Spectroscopy (SXR) beamline of the Australian Synchrotron, part of ANSTO. The authors also acknowledge the National Energy Research Scientific Computing Center (NERSC) provide supercomputer resources to this work for the DFT calculations, and RSoXS fits. The authors also acknowledge Dr. Thomas Ferron, Dr. Cheng Wang, and Dr. Kas Andrle form the Advanced Light Source of Lawrence Berkeley National Laboratory for their training and guidance in data collection, reduction, and fitting analysis. The authors also thank Dr. Arda Gozen from the School of Mechanical and Materials Engineering of Washington State University for AFM support on measuring soft nanoscale films. Fable 5 from Anthropic was used for minor copy editing and code generation under the ALS computing group's team plan. The authors also thank Dr. Daniel Dougherty from North Carolina State University and Dr. Dean DeLongchamp from the National Institute of Standards and Technology for their reading and consultation of the manuscript.

**Author Contributions:** HH and BAC conceived the study. BAC oversaw and directed all aspects of the study. HH performed RSoXS measurements, DFT calculations, developed and performed the structural and optical modeling, and wrote the manuscript. FW deposited samples measured with RSoXS and performed and measured AFM topology. EB deposited a separate sample, collected, and analyzed non-resonant XRR. SA performed and analyzed VASE. TM collected and analyzed GIWAXS and measured AFM topography. VM assisted with DFT calculations and clustering. LT collected AR-NEXAFS and OA processed and analyzed the AR-NEXAFS.

**Funding:** This work was supported by the Department of Energy Early Career Research Program under Grant DE-SC0017923. HH was supported in part by an ALS Doctoral Fellowship in Residence. Funding support for OA was provided by Prince Sattam bin Abdulaziz University, Saudi Arabia.

**Competing interests:** The authors declare no competing interests.

**Supplemental information:** The online version contains supplementary material

**Correspondence and requests for materials** should be addressed to Brian A. Collins

# Supporting Information

Harlan Heilman[1], Emily Biegler[1], Sushil Adhikari[1], Victor Murcia[2], Tanner Melody[1], Fred Woodall[1], Lars Thompson[3], Obaid Alqahtani[2,4], Brian A. Collins[1,2*]

[1]*Department of Physics and Astronomy, Washington State University, Pullman, Washington 99164, USA*

[2]*Department of Materials Science and Engineering, Washington State University, Pullman, Washington 99164, USA*

[3]*Australian Synchrotron, ANSTO, Clayton, VIC, Australia*

[4]*Department of Physics, Prince Satam bin Abdulaziz University, Alkharj 11942, KSA*

## 1. Film pre-characterization

### *1.1. Variable angle spectroscopic ellipsometry*

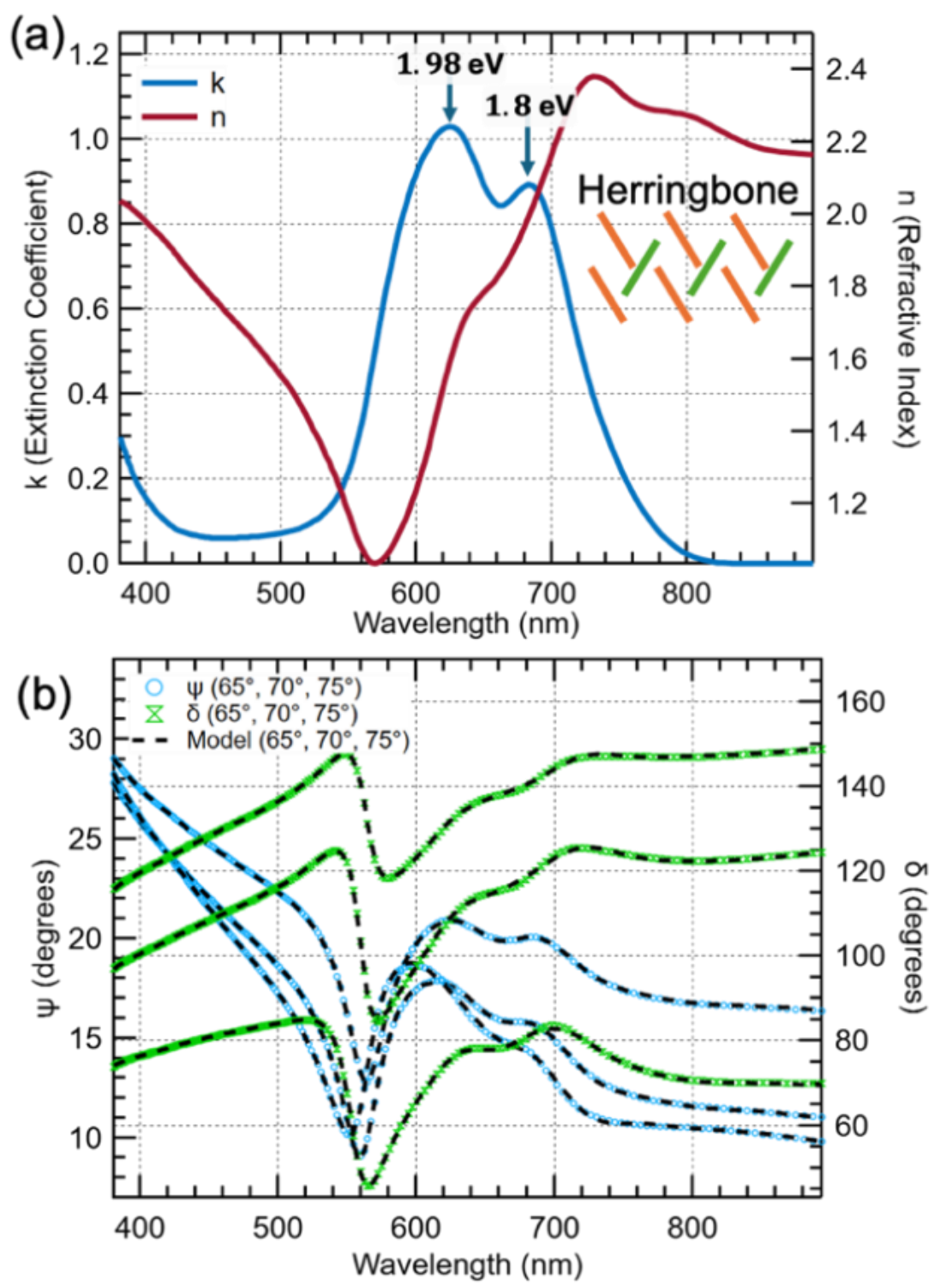

FIG. S1 VASE for ZnPc films. **(a)** Optical constants derived from the VASE fit, highlighted is the Q-band splitting at 625 nm (1.98 eV) and 690 nm (1.80 eV) indicative of herringbone phase ZnPc packing motifs. **(b)** Fit results for each incidence.

The extinction coefficient exhibits a split Q band at 1.98 and 1.80 eV characteristic of Davydov coupling between the translationally inequivalent molecules of the herringbone packed motif. That the higher energy peak is more intense, suggests the edge on herringbone ($\alpha$ family) polymorph previously reported for as-deposited ZnPc in both UV-Vis and ellipsometry derived optical constants.[52,66] ZnPc layer thickness is recorded as $t_{ZnPc} = 21.4(5)$ nm with a MSE of 1.433, consistent with the QCM report during deposition.

### *1.2. Grazing incidence wide angle X-ray scattering*

The GIWAXS texture corroborates the polymorph and orientation assignments assigned though ellipsometric optical constants. The out-of-plane (200) reflection at $d = 12.6$ Å together with the in-plane (002) and $\pi - \pi$ stacking reflections place the film in agreement with edge-on textured ZnPc on weakly interacting oxides.[18,66] The (200) spacing matches the relaxed $\gamma$-ZnPc (herringbone / $\alpha$ −II family) value of 12.67 Å rather than the strained early growth spacing observed in the initial layers of growth.[18]

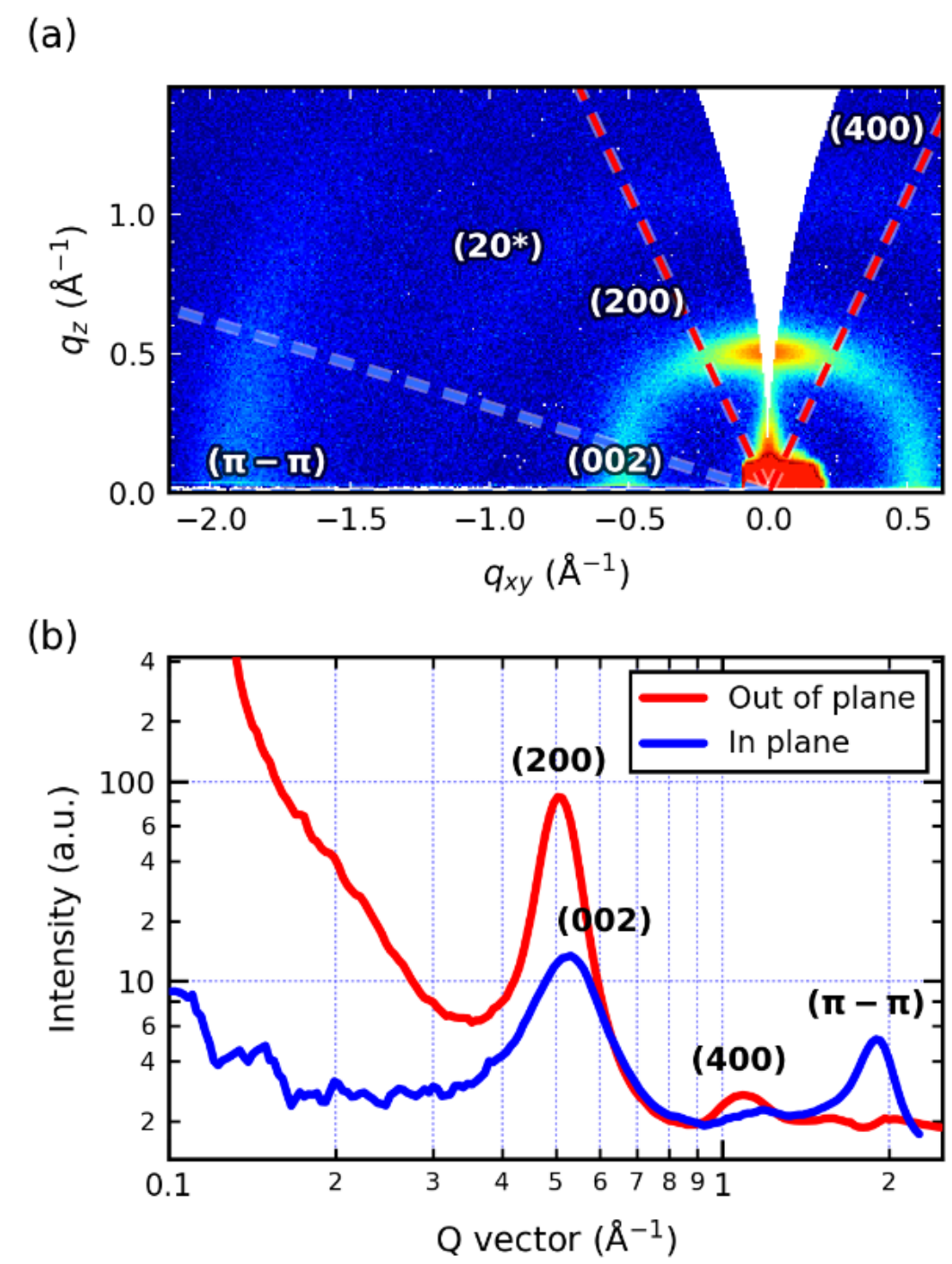

FIG. S2 GIXRD for ZnPc films. **(a)** GIWAXS image of ZnPc deposited on native-oxide silicon substrate. **(b)** 1D reduction showing the out-of-plane (200) peak at 0.50 Å$^{-1}$ (d-spacing 12.6 Å), it's broadened second order (400) near 1Å$^{-1}$, the in-plane (002) peak at 0.53 Å$^{-1}$ (11.8 Å) as well as a mixed index $\pi - \pi$ stacking peak at 1.89 Å$^{-1}$ (3.3 Å). Weak arks near $\pm 45°$ from the substrate normal, makers (20*) in **(a)**, are tentatively assigned to a mixed-index reflection but are too broad azimuthally for a definitive index.

### *1.3. Atomic force microscopy*

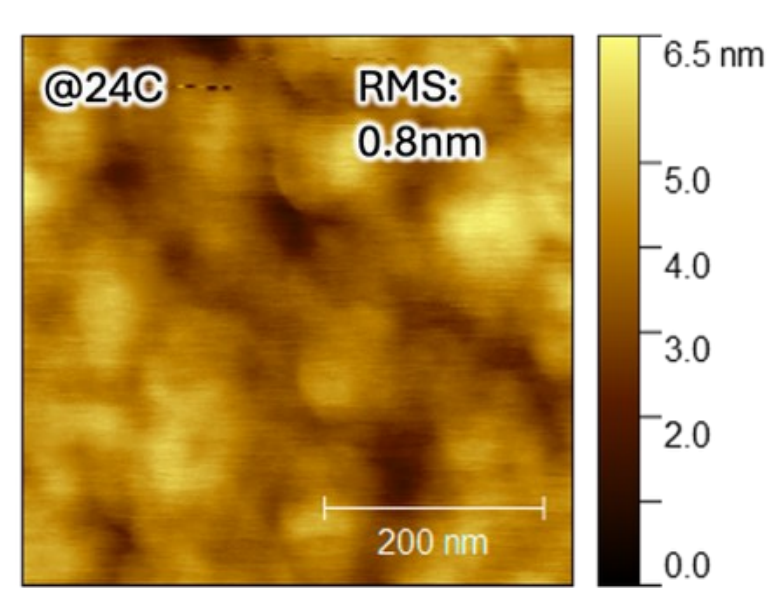

FIG. S3 AFM height map of ZnPc sample from the same batch used for RSOXS measurement. RMS roughness measured as 0.8 nm.

Atomic force microscopy (AFM) height maps were acquired on a Bruker Dimension Icon in tapping mode with a TESPA-V2 SAMPLE probe. Scans where processed using Gwyddion[55] to extract root-mean-square (RMS) roughness values. Film morphologies and RMS roughness values were used to pre-characterize film surface roughness and morphology reflectivity measurements (**Fig S3**).

*1.4. Effect or growth parameters for Copper (II) Iodide patterned films*

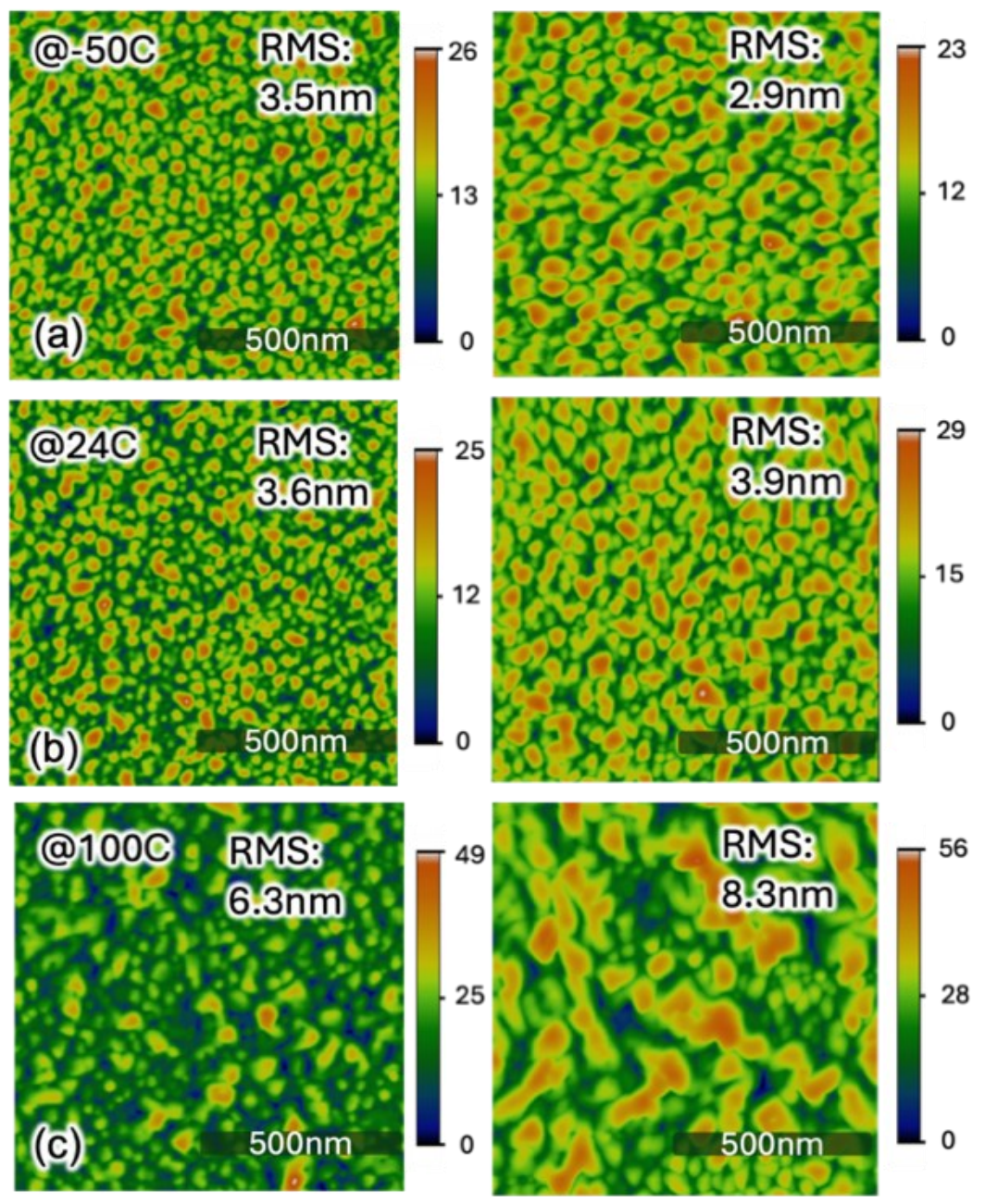

FIG. S4 AFM height map of ZnPc under different deposition conditions. **(a-c)** ZnPc deposited on a 3 nm CuI patterning layer as a function of deposition rate (left .15 $Å/s$, right $.5\ Å/s$) and substrate temperature **(a)** -50 C, room temperature **(b)** 24 C, and **(c)** 100 C.

A film series was prepared for ZnPc grown on a 3 nm CuI patterning layer to identify deposition conditions to make suitable RSoXS samples. Samples were grown to a target of 20 nm as measured via QCM and validated through spectroscopic ellipsometry. AFM height maps were performed for the series and displayed below.

The surface texture for these films shows dense, laterally uniform grains with significant RMS roughness across the series. Increasing temperature leads to the formation of larger grains with more dispersity amongst their size. At 100C the morphology is most dependent on deposition rate, with the higher rate resulting in elongated domains and a larger RMS roughness. Across all conditions, film heights vary by more than the approximate layer thickness (> 20 nm) suggesting that ZnPc does not form fully wetted and continuous layers when deposited on CuI, in agreement with prior work in device applications.[67] As such we surmise that when deposited on CuI, layers grow under the Volmer-Weber (island) process, developing towers of ZnPc as opposed to continuous layer formation. Reflectivity was attempted but could not be reduced to a reliable specular dataset due to the dominant off specular signal, consistent with incomplete surface coverage

*1.5. NEXAFS Spectroscopy*

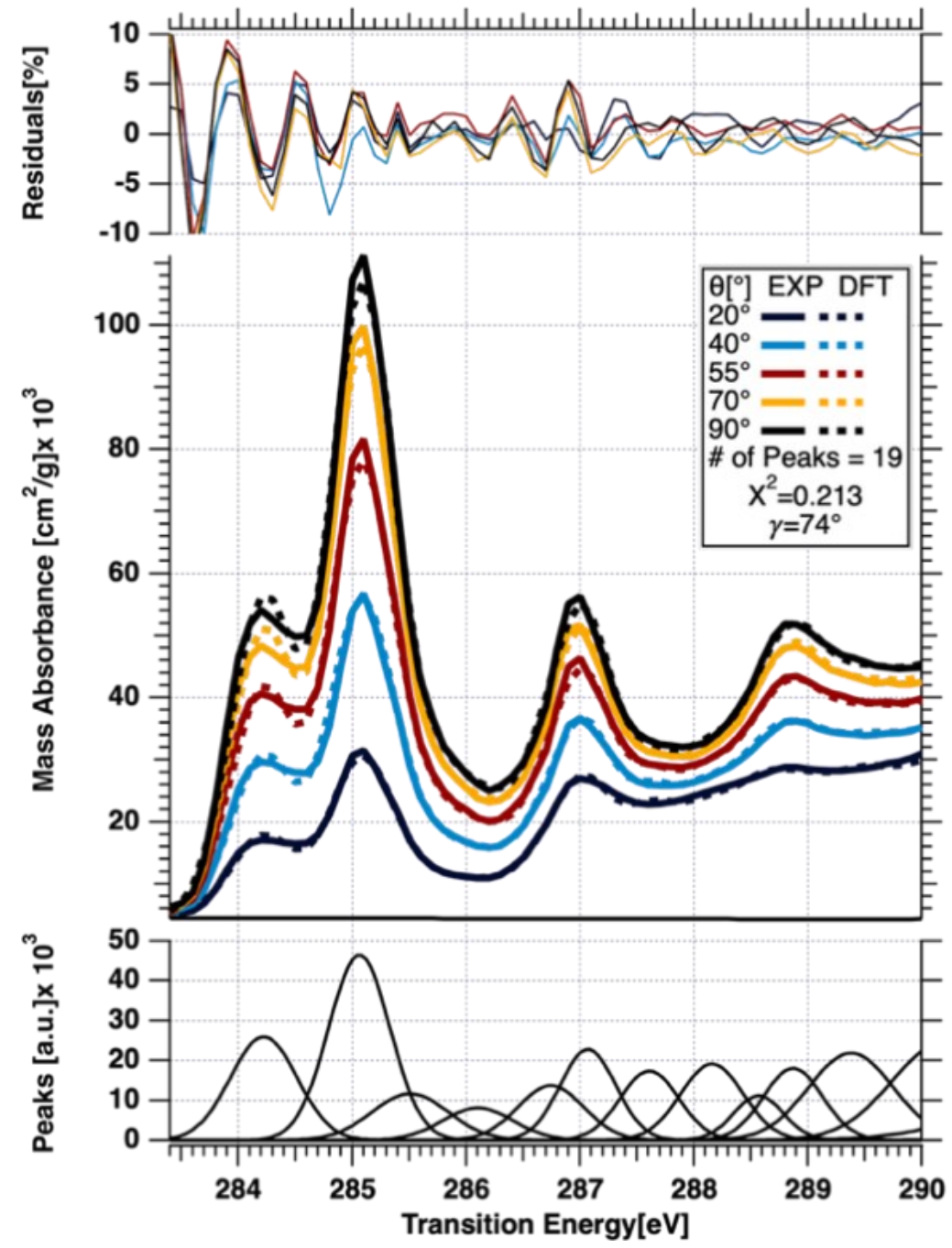

FIG. S5 Comparison of refined clustered optical model with experimental NEXAFS spectroscopy. Tilt angle molecular tilt angle is 74°determined though the refinement of clustered DFT.

NEXAFS spectroscopies were collected at the Soft X-ray Spectroscopy (SXR) beamline of the Australian Synchrotron, part of ANSTO in total electron yield mode on Si wafers. Samples were prepared under similar conditions as for RSoXS measurements yet were collected via a separate batch. Spectra were collected at five sample alignments with respect to the incident electric field polarization vector with respect to the surface normal $(20°, 40°, 55°, 70°, \& 90°)$. NEXAFS was collected over the energy range of 270-350 eV and normalized to the energy dependent intensity fluctuation and optical loss via a direct photodiode measurement, and Au mesh drain current. Mass absorption was computed via normalization in the pre and post edge regions to the bare atom mass absorption of ZnPc as extracted from the CXRO Henke atomic scattering factors.[57]

Molecular tilt angle was extracted via the DFT cluster and refinement process. As the final step of the refinement process, the molecular polar tilt angle $\gamma$ is fit using the TDM cluster orientations prescribed by DFT providing a measure of the molecular orientation. This fit produces a reduces $\chi^2$ goodness of fit statistic of .213 indicating very good agreement between the model and experimental data. The molecular tilt angle of 74° agrees with all other characterizations of the ZnPc face on herringbone polymorph described in this study and is taken as the coarse-grained molecular orientation within the films measured under RSoXS.

## 2. Optical Tensor Construction

The optical tensor is defined under the convention $n(E) = 1 - \delta(E) + i\beta(E)$. From experimental NEXAFS the imaginary component is computed directly form the mass absorption spectra:

$$\beta(E) = \mu\rho/2k \qquad (S1)$$

where $\mu$ is the mass absorption, $\rho$ is the mass density, and $k$ the wave number.

The NEXAFS spectra is scaled to agree with the bare atom mass absorption as tabulated by the CXRO with a density of one.[68] This constructs an index of refraction specific volume that we use to evaluate the index of refraction for a given material density. The real part of the refractive index $\delta(E)$ is computed via the Kramers-Kronig calculation using the KKcalc package.[69]

The experimental tensor components are constructed following the established empirical method.[27,49] From clustering, the primary $\pi^*$ TDM cluster rests at $.01°$ from the molecular normal. The NEXAFS derived dichroism across the carbon edge is computed by taking the average difference spectra between all pairs of incident angles as a function of energy. This constructs an experimental dichroism that is used to augment the isotropic $\beta_{iso}(E)$ measured at the magic angle to recover the refractive tensor components from the molecular reference frame, i.e.,

$$\beta_{zz}(E) = \beta_{iso}(E) + A(\theta)\Delta\beta(E) \qquad (S2)$$
$$\beta_{xx}(E) = \beta_{iso}(E) - B(\theta)\Delta\beta(E) \qquad (S3)$$

The coefficients $A(\theta)$ and $B(\theta)$ are defined to match the condition in Stöhr, modified to accommodate the off axis TDM polar angle $\theta$ associated with the of the primary transition.

$$A(\theta) = (3\cos^2\theta - 1)\,\beta_{iso}(E_\pi)/\Delta\beta(E_\pi) \qquad (S4)$$
$$B(\theta) = (1 - \cos^2\theta/2)\beta_{iso}(E_\pi)/\Delta\beta(E_\pi) \qquad (S5)$$

The DFT tensor is instead defined using the directly calculated tensor components for each TDM cluster, resulting in the comparison of Fig 1, requiring no auxiliary analyses.

## 3. DFT Methods

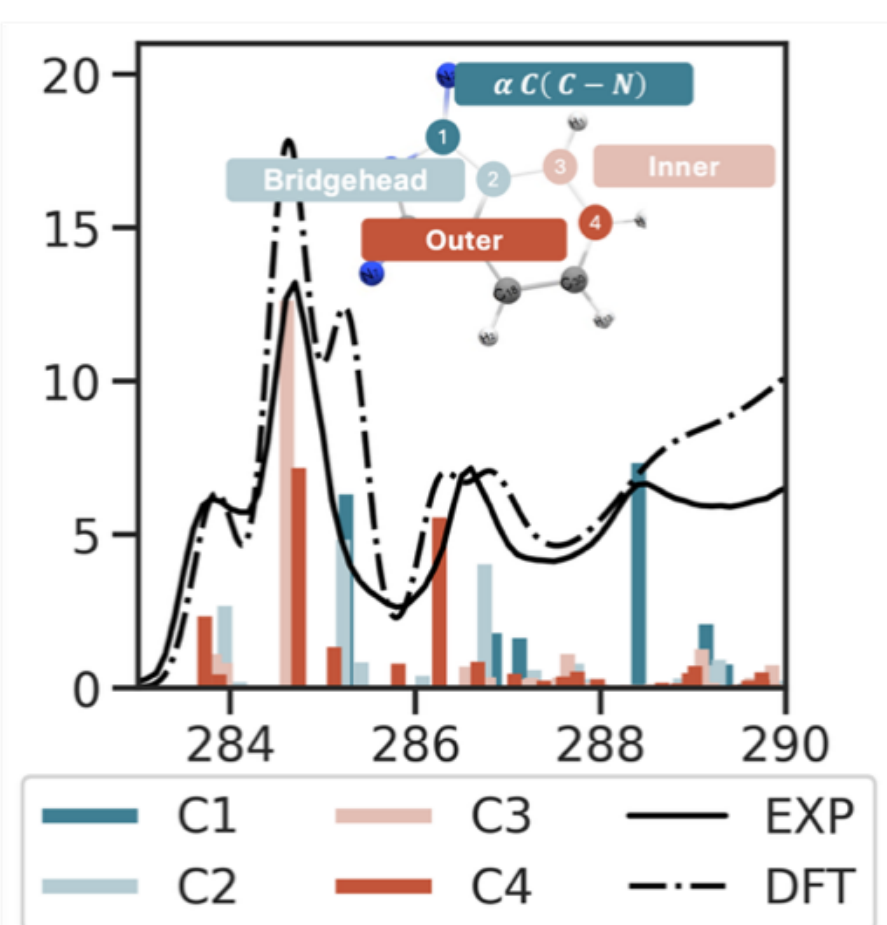

FIG. S6 DFT calculated NEXAFS spectroscopy for ZnPc computed from the four highlighted excitation centers vs transition energy, overlapped with experimental NEXAFS from the SXR beamline. Poles indicate individual contributing oscillators.

A time independent transition potential calculation using the StoBe library was performed on a single molecule of ZnPc. A HCH approach sets the manifold of transition energies, along with each TDM and oscillator strength, while the $\Delta KS$ method is used to fix the absolute energy associated with each transition. For details see our previous work.[33] The molecular geometry was extracted from a crystallographic structure file (.cif) as it produced results with a higher initial agreement to experiment compared to thermodynamically relaxed structures, or those relaxed from molecular force fields.

From this structure, four chemically distinct carbon atoms (labeled in **Fig 1 & S6**) set the primary excitation centers for the molecule. Calculations using the eight carbons of the entire outer molecular wing, along with the 16 for half of the molecule, were tried yet offered no improved accuracy and only increased dimensionality and complexity of the calculation results. Here we demonstrate that a reduced set of excitation reproduces the entire set of transitions possible for molecular ZnPc, and though clustering provides an optical model that is useful for scattering analysis.

### *3.1. Clustering and Refinement*

Clustering and refinement were performed using the Igor Pro implementation of the hierarchical agglomerative clustering to extract a set of TDM clusters as the initial building blocks for the optical model. Oscillator strength threshold and Overlap percentage hyperparameters were explored using an evenly spaced support via

$$OS \in [0, 10] \;\&\; OVP \in [25,75]$$

Hyperparameter pairs were evaluated by reduced chi square to experimental NEXAFS and well-behaved parameter pairs where secondarily evaluated by their difference form the original DFT, ensuring the preservation of the DFT calculated information. Here we chose the best fit clustering parameters of $OS = .25$ and $OVP = 52.5$. In general, we find good clustering and convergence for all $OVP \in [55,50]$, with significantly worse performance outside of these bounds. Moreover, we find that $OS \in [.1,8]$ performed similarly, yet parameter agreement with originating DFT degrades rapid for larger $OS$, while the refinement never converse for $OS < .1$.

## 4. Optical model parameterization

Molecular order within the film is parameterized by the molecular tilt angle $\gamma$ of the molecular surface normal from the film surface normal. For uniaxial film, this is the polar cone angle. We begin with the single molecule refined DFT optical model. This construction assumes the single molecule DFT is representative of the electronic states available to a film of the material, ignoring the creation of new multi molecule states, that could be caused by variations to molecular packing or orientation alignment. However, as the clustered DFT model agrees with the experimentally derived optical model, we surmise that these effects are not dominant in this system.

We start with the uniaxial molecular optical tensor $n^m$ constructed through clustering,

$$n^m(E) = \begin{pmatrix} n_{xx} & 0 & 0 \\ 0 & n_{xx} & 0 \\ 0 & 0 & n_{zz} \end{pmatrix} \qquad (S6)$$

where $n_{xx}$ is the in plane or ordinary component, and $n_{zz}$ is the out of plane or extraordinary component. The molecule is then tilted by an arbitrary polar angle $\gamma$ about the $x$ or $y$-axis, the choice being arbitrary, before computing the uniform azimuthal average over $\phi \in [0,2\pi]$, i.e.,

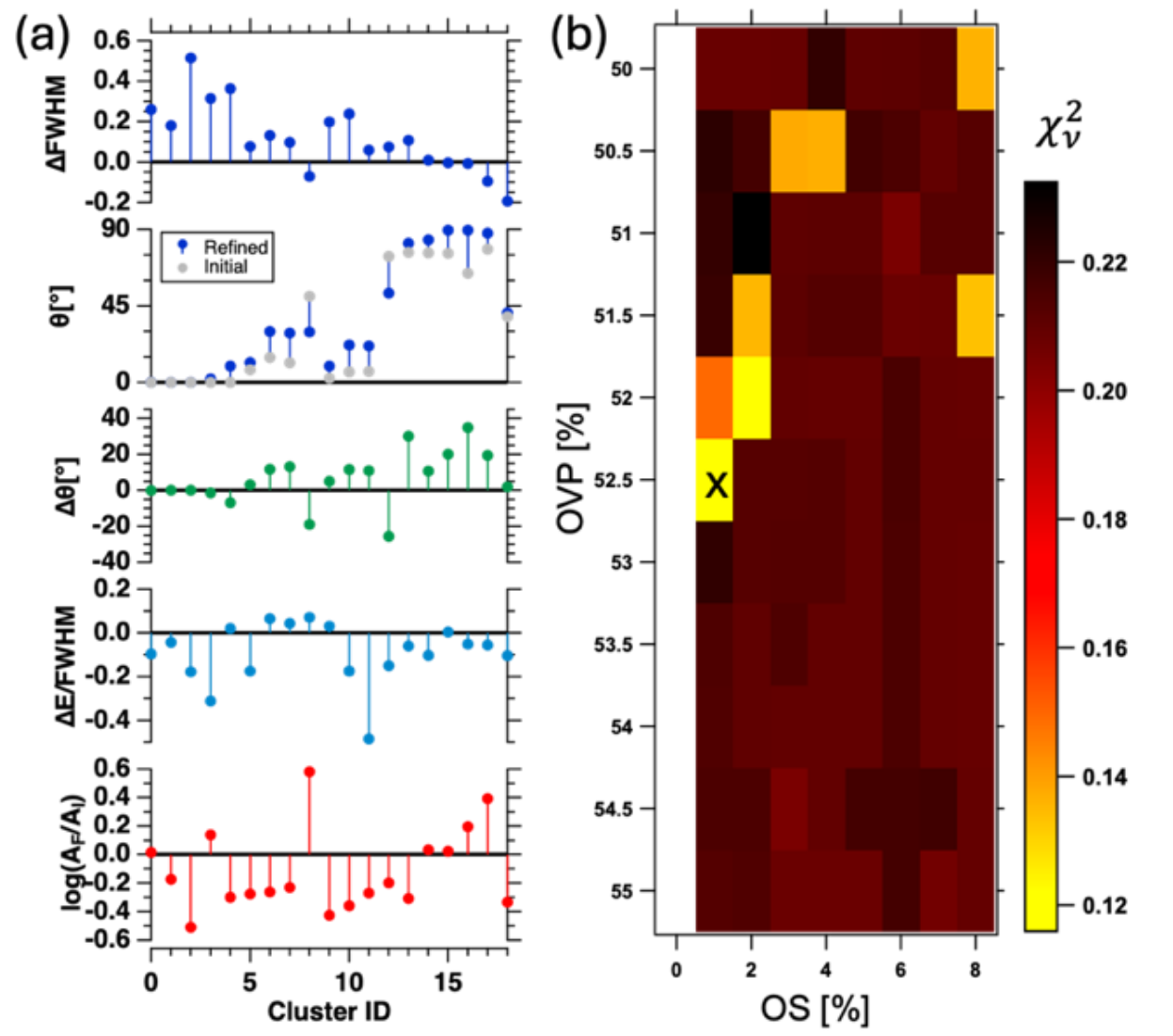

FIG. S7 Clustering and refinement analysis. (a) Change in parameters though the refinement process for the chosen set of hyperparameters. (b) OVP % and OS % hyperparameter optimization map

$$n^f(E,\gamma) = \langle R^T(\phi,\gamma) n^m(E) R(\phi,\gamma)\rangle_\phi \qquad (S7)$$

where $\phi$ is the dummy azimuthal orientation variable, and $R(\phi,\gamma)$ the product $R_z(\phi)R_{x/y}(\gamma)$. Because the material and refined molecular tensor are azimuthally symmetric, the off diagonal components vanish by inspection and, with a density scale factor, the ZnPc optical tensor reduces to,

$$n_{xx}^f(E;\gamma,\rho) = \frac{\rho}{2}\left(n_{xx}^m(1+\cos^2\gamma) + n_{zz}^m \sin^2\gamma\right) \qquad (S8)$$

$$n_{zz}^f(E;\gamma,\rho) = \rho\left(n_{xx}^m \sin^2\gamma + n_{zz}^m \cos^2\gamma\right). \qquad (S9)$$

To validate these results, we find the isotropic condition is $\gamma = \arcsin\sqrt{2/3} \approx 54.7°$ the so-called magic angle, in agreement with Stöhr.[48]

## 5. RSoXS Experimental and Analysis Methods

### *5.1. RSoXS Collection*

RSoXS measurements were performed at advanced light source beamline 11.0.1.2 following the standard procedures established by Ferron et. al. Using the scatter slits and higher order suppressor optical elements, the direct beam intensity was attenuated to a reasonable level for direct measurement on the photodiode detector. In this configuration, the beam was aligned using the in-chamber photodiode detector via the following. A knife edge experiment is used to align the sample z-motor position. Then from this z position, the sample theta motor is offset to 1-2 degrees. The detector theta motor is then scanned over a 4-degree window around the corresponding 2theta position. By recursively performing sample-z-and-theta alignment scans the sample can be aligned to the center of the goniometer axis of rotation to within .01 deg. Note that motor issues, beam drift, and tilt along the sample plate, the axis alignment must be recalibrated during extended runs, especially after changing polarization or substantially changing beamline energy.

Experimental data was collected starting in the s-pol configuration, beginning at 250 eV and working to the resonant energies realigning the axis of rotation after every 10 energies. Between polarizations, the alignment was recalibrated after waiting for 30 min for the new heat load to affect the optics. Measurements were performed over a 12-hour shift with an energy calibration performed using the transmission NEXAFS of polystyrene, at the start and end of the shift showing no energy drift throughout the experiment duration.

### *5.2. Reflectivity Reduction*

Reflectivity measurements are reduced from 2D images by first locating the specular reflected beam on the area detector and integrating the intensity within a suitable region. A background signal is removed from this intensity using a region of the detector far from the specular beam spot. This background region is collected along the same row as the beam spot to account for systematic CCD readout noise. The measured reflectance is then computed from the specular reflected intensity by dividing by the incident $I_0$ intensity collected at the start of each energy scan.

High dynamic range sensitivity was achieved by slicing the measurement q-range into subsets with overlapping points where the reflected intensity could be well captured on the two orders of magnitude of dynamic range allotted by the CCD detector. This is done by using the higher order suppressor optical element and increasing the exposure time. Using the overlapping points, an effective $I_0$ intensity can be determined for each subset used in the calculation of reflectance for long q-range scans.

### *5.3. Error Approximations*

Each frame of a reflectivity profile is inherently a photon counting measurement. The raw intensity in each detector pixel follows a poissonian distribution to first approximation, giving a per-pixel error equal to $\sqrt{N}$. Systematic noise is also present, originating from the dark current, readout noise, and stray light. Readout noise is reduced by first trimming a 10-pixel border around the images and then performing a row-by-row subtraction using the average counts from the first 10 pixels in every row of the detector image. This flattens the thermal background caused by the readout process. The stray light and dark current can then be modeled and accounted for using the dark region of the detector as described above. Non-poissonian noise is characterized using a Fano factor defined as the ratio of the observed variance in $I_0$ and effective $I_0$ measurements to the hypothesized poissonian variance at the same intensity level. This effectively scales the raw poissonian noise in each q subset.

The final source of error comes in the discrimination between specular and off specular reflectance. Thin film surface scattering and roughness effects smear the reflected beam over a finite width of pixels. This effect becomes more pronounced at higher q-values. To account for this error a q-dependent uncertainty envelope is applied to the propagated uncertainties from data reductions producing a higher uncertainty in the higher q regions of spectra.

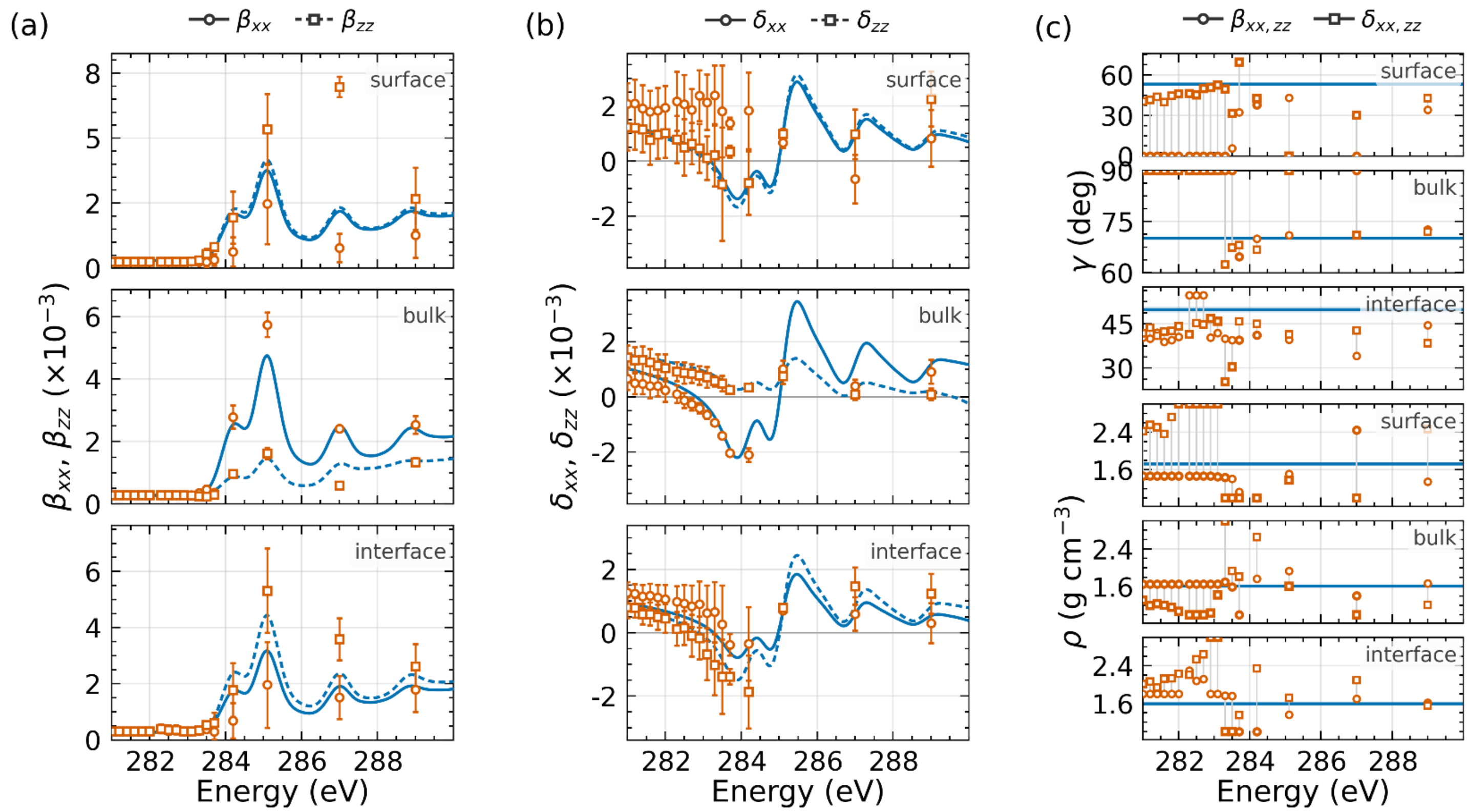

FIG. S8 Comparison in final best fit optical constants revealed through the Free optical tensor and dft models. **(a)** Imaginary tensor components $\beta$, **(b)** real tensor components $\delta$. The extracted orientation and density are compared in **(c)** and **(d)** where the directly refined orientation from the DFT model (traces) is compared with the per energy orientation computed from a secondary refinement to the full tensor. The density is determined by the magnitude of the trace of the optical tensor and showcases a much lower inter energy variance. The orientation that sets dichroism and birefringence varies much strongly between energies and amongst q-ranges.

*5.4. RSoXS Analysis Methods*

Our fitting methodology has been largely updated from the procedures outlined in Ferron et. al.[27] The model was constructed containing three layers of ZnPc (Surface, Bulk, Interface), on top of a thin SiO2 native oxide layer, followed by the Si substrate. Each ZnPc layer is modeled using a uniaxial optical tensor whose derivation is described in the methods above. The reflection and transmission coefficients for the stack are computed for each energy and scattering vector via the 4x4 transfer matrix formalism where boundary condition matching at each of the five interfaces sets the reflection and transmission coefficients for each layer, while interfacial roughness manifests as an exponential attenuation to account for the off specular diffuse scattering from a faceted interface. Here we model interfacial roughness in the Nèvot-Croce (NC) sense as is standard for reflectivity analyses.[62,70] This sets the parameterization of the physical model structure as thickness and roughness for the surface, bulk, and interface layers of ZnPc, along the native SiO2 layer for a combined eight parameters to set the full structure.

The optical constants for each layer are set depending on the optical model and differ between the free tensor and DFT parameterization. Consistent across both are the non-resonant material slabs for the vacuum and silicon fronting media, along with the silicon dioxide layer. Each are set with tabulated bare atom optical constants, parameterized by the energy independent material density that we allow to vary subtly from the tabulated values to accommodate differences in stoichiometry, but fixed to a feasible region of (2, 2.8) g/cm$^3$. The optics for free tensor ZnPc are parameterized as birefringence, dichroism, isotropic $\delta_{iso}$, and isotropic $\beta_{iso}$ each constrained to a reasonable range dictated by the original optical model and following the choices by Ferron et. al. resulting in 252 optical tensor components for the present work. The DFT optics are set energy independently via a molecular orientation and density per layer, along with an energy offset term common across all layers and accounting for the energy calibration differences between instruments resulting in just seven parameters to describe the three ZnPc layers. Thus, the free tensor parameterization uses 266 (8+2+252) parameters to describe the material, while the DFT model uses just 17 (8+2+7) to describe the same structure, over an order of magnitude in reduced complexity.

The free tensor parameterization is the most general model and differs from the best analyses by Ferron et. al.[27] In prior works, a dichroic ratio is used to set the ratio between the birefringence and dichroism for the material allowing for a reduction of one degree of freedom per energy in the model. However, correctly computing this requires the creation of the energy dependent optical tensor wile resulting in only a 3/4 reduction in optical parameters still requiring the fitting of the birefringence per energy. This fails to enforce cross energy consistency checks such as the Kramers-Kronig relations or the consistency in molecular orientation between energies. As a result, in our comparisons, we take the full tensor complexity for our Free tensor model.

The reflectivity model calculating reflectivity from the material stack is augmented to include several experimental correction terms that account for beam and motor drifts, along with instrument miscalibration during the measurement. This includes a normalization correction scale factor and an angle of incidence offset applied to each energy and polarization, following the analysis of Ferron et. al.[27] Combined this adds in four correction terms per energy (84 in this work) accounting for a majority of the fitted parameters in our DFT constrained optical model. This brings the total model parameters to 101 for the DFT model and 350 for the Free tensor model.

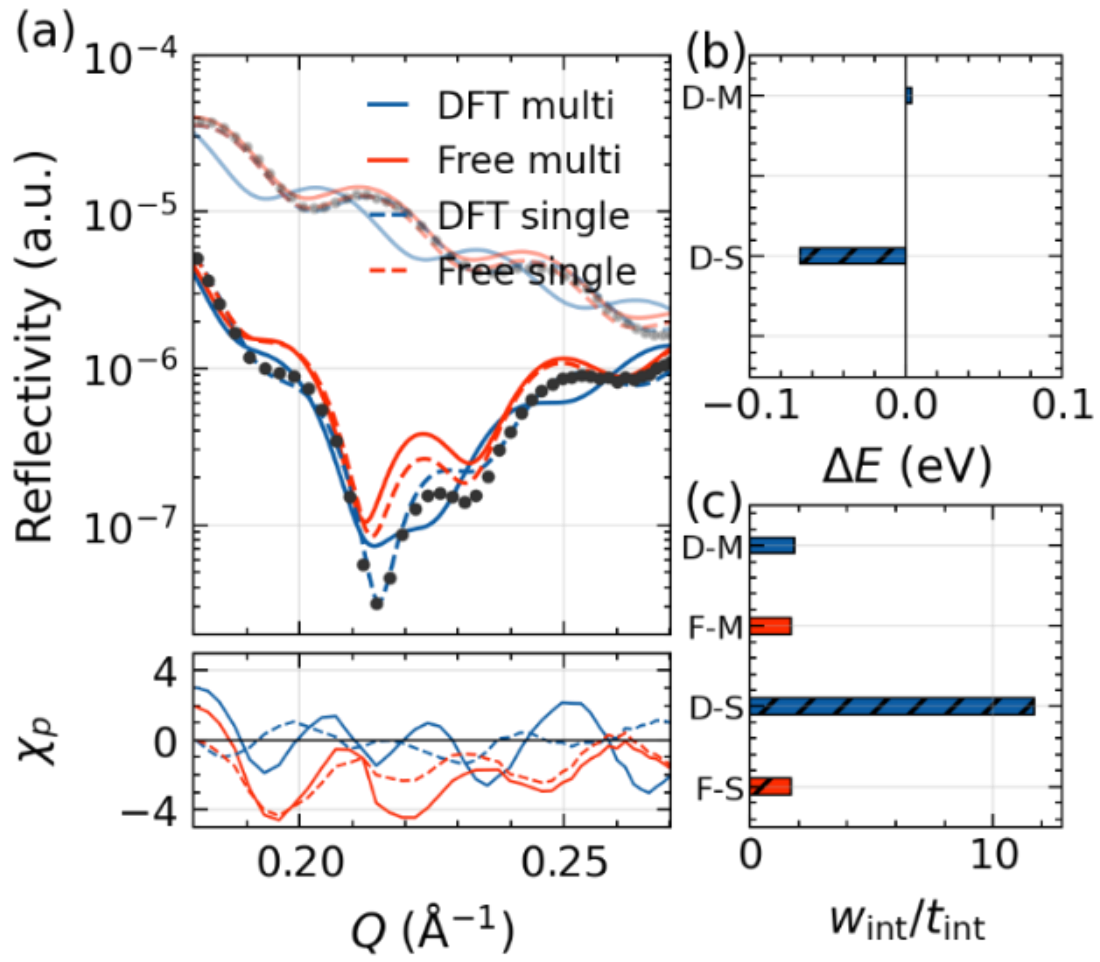

FIG. S9 Comparison of single energy reflectivity fits at the diagnostic energy (283.7 eV) against the multi energy refinement. **(a)** Direct fit comparison and residuals showcasing that the DFT single energy can out-perform other fit strategies within Brewster's angle. **(b)** Energy offset parameter results used in the fits compared between the DFT multi energy (D-M) and single energy (D-S) fits indicating that the single energy fit must take an extreme -.1 eV offset, to a non-absorbing energy. **(c)** Interfacial width $w = \sqrt{2\pi}\sigma_r$ of the interface layer showcasing that the D-S fit is >10x the total thickness of this buried layer.

For those seeking to follow this analysis, experimental correction parameters can often be critical to producing a good fit. However, it is important to note that without proper bounds on a per energy basis, experiment correction parameters easily become cross-correlated with both the structure and optics leading to non-physical fit results. We recommend only adding these offset parameters if the first fringe is challenging to reproduce using physically realistic structural and optical parameters. Reasonable bounds on the normalization correction term are (.8, 1.2) with exceptions for datasets with visible outliers in the $I_0$ measurement points. Angle of incidence may vary more widely from zero, and manual observations of the data should be used to set a feasibility region of this parameter. Here we use bounds of (-.1, .1) due to the measurable alignment drift across energies. Additionally, as ALS 11.0.1.2 and other future RSoXS capable instruments constantly improve their configurations, these ranges should be discussed with the beamline scientist for the instrument for the latest insights.

*5.5. Fit Results*

TABLE SI Best fit structural properties of the ZnPc film extracted using the Free tensor optical model. Errors are reported as $2\sigma$ from the covariance matrix.

| Layer | $t$ (Å) | $\sigma_r$ (Å) |
|---|---|---|
| **Surf** | 8.8(4) | 7.21(28) |
| **Bulk** | 174.4(4) | 10.43(31) |
| **Int.** | 14.48(24) | 8.39(15) |

TABLE SII Best fit structural properties of the ZnPc extracted using the ML-DFT slab optical model. Errors are reported as $2\sigma$ from the covariance matrix.

| | $t$ (Å) | $\sigma_r$ (Å) | $\gamma(\circ)$ | $\rho(g/cm^3)$ |
|---|---|---|---|---|
| **Surf.** | 8.933(4) | 7.056(1) | 52.21(7) | 1.8(3) |
| **Bulk** | 179.95(3) | 13.136(2) | 69.54(1) | 1.638(3) |
| **Int.** | 10.705(2) | 7.892(5) | 47.24(9) | 1.0(6) |

### 6. Orientation gradient theoretical foundation

We formulate our gradient model from a Ginzburg-Landau framework of second order phase transitions, where we treat the molecular orientation and density as order parameters (not to be confused with nematic order parameters $S(\gamma)$ or $P(\gamma)$).[71,72] Along the one dimensional depth axis $z$ of a film of thickness $L$, we take the order parameters as the deviation in the density $\Delta\rho(z) = \rho(z) - \rho_b$ and orientation $\Delta\gamma(z) = \gamma(z) - \gamma_b(z)$ from the bulk. The total free energy of the film is described by the functional

$$F[\Delta\rho, \Delta\gamma] = \int_0^L f(\Delta\rho, \Delta\gamma) + \frac{1}{2}c_\rho(\Delta\rho')^2 + \frac{1}{2}c_\gamma(\Delta\gamma')^2 dz + F_v[\Delta\rho(0), \Delta\gamma(0)] + F_{Si}[\Delta\rho(L), \Delta\gamma(L)] \qquad (S10)$$

Here $f(\Delta\rho, \Delta\gamma)$ is the local bulk free energy density, and the terms proportional to $c_\rho$ and $c_\gamma$ are the gradient energy penalties that oppose abrupt variations within the film. The interface terms $F_v$ and $F_{Si}$ account for the anchoring potentials at the interface boundaries. Expanding $f(\Delta\rho, \Delta\gamma)$ to quadratic order about the bulk minimum,

$$f(\Delta\rho, \Delta\gamma) \approx f_0 + \frac{1}{2}a_\gamma\Delta\gamma^2 + \frac{1}{2}a_\rho\Delta\rho^2 + b\Delta\rho\Delta\gamma. \qquad (S11)$$

where $a_\rho$ and $a_\gamma$ are the stability coefficients of the bulk phase, and $b$ the bilinear coupling term between the density and orientation fields.

Minimizing the total variation $\delta F = 0$ and integrating the gradient terms by parts yields a system of coupled second-order differential equations for the film interior,

$$c_\rho\Delta\rho'' = a_\rho\Delta\rho + b\Delta\gamma \qquad (S12)$$
$$c_\gamma\Delta\gamma'' = a_\gamma\Delta\gamma + b\Delta\rho \qquad (S13)$$

with boundaries dictated by the surface free energies. To leading order in the interface fields $h$, the boundary variation yields

Robin boundary conditions, where the gradient of the order parameter is proportional to the surface field, i.e.

$$c_\rho \Delta\rho'(0) = -h_\rho^v \qquad c_\rho \Delta\rho'(L) = h_\rho^{Si} \qquad (S14)$$

and with similar form for $\Delta\gamma$. In the limit of strong anchoring, the surface fields $h$ are much stronger than the bulk stiffness coefficients $c/\tau$, the system enters the Dirichlet regime. Here each interface potentials set the order parameters to fixed interfacial values, $\Delta\rho(0) = \Delta\rho_v$ and $\Delta\rho(L) = \Delta\rho_{Si}$. Within this regime the interfaces impose an orientational state regardless of the bulk interaction, a feasible state given the known different growth modes for ZnPc films at the substrate and surface.[18,46]

This system of equations may be decoupled into a normal modes problem, under the coordinate transform $\widetilde{\Delta\rho} = \sqrt{c_\rho}\Delta\rho$ and similar transform for $\Delta\gamma$. Given that the bulk represents the stable minimum, the general solution is restricted to the linear combination of exponentials decaying from the boundaries with coherence lengths $\tau_\pm$ derived from the eigenvalues of the coupled system. When the orientation and density coupling is nonzero, these coherence lengths are shared between the order parameters. As each interface bears its unique interactions that dictate this coherence length, we assume that each interface is dominated by a single mode. The interface gradient model used in the main text satisfies these conditions,

$$\Delta\rho(z) = \Delta\rho_v \exp(-z/\tau_{vac}) + \Delta\rho_{Si} \exp(-(L-z)/\tau_{Si}) \quad (S15)$$
$$\Delta\gamma(z) = \Delta\gamma_v \exp(-z/\tau_{vac}) + \Delta\gamma_{Si} \exp(-(L-z)/\tau_{Si}) \quad (S16)$$

While this model is congruent with the continuous reorientation under an evolving growth front, it remains an empirical approximation of the underlying interfacial physics. A more rigorous determination of the anchoring potentials would require a detailed investigation of the boundary conditions, specifically whether the system is governed by Dirichlet or Robin boundaries as this determines the relationship between exponential amplitudes and the interfacial potentials. Furthermore, this model assumes a continuous structural relaxation; however, it is possible that the molecular realignment across the interface occurs under a different mechanism than described here requiring a different theoretical basis.

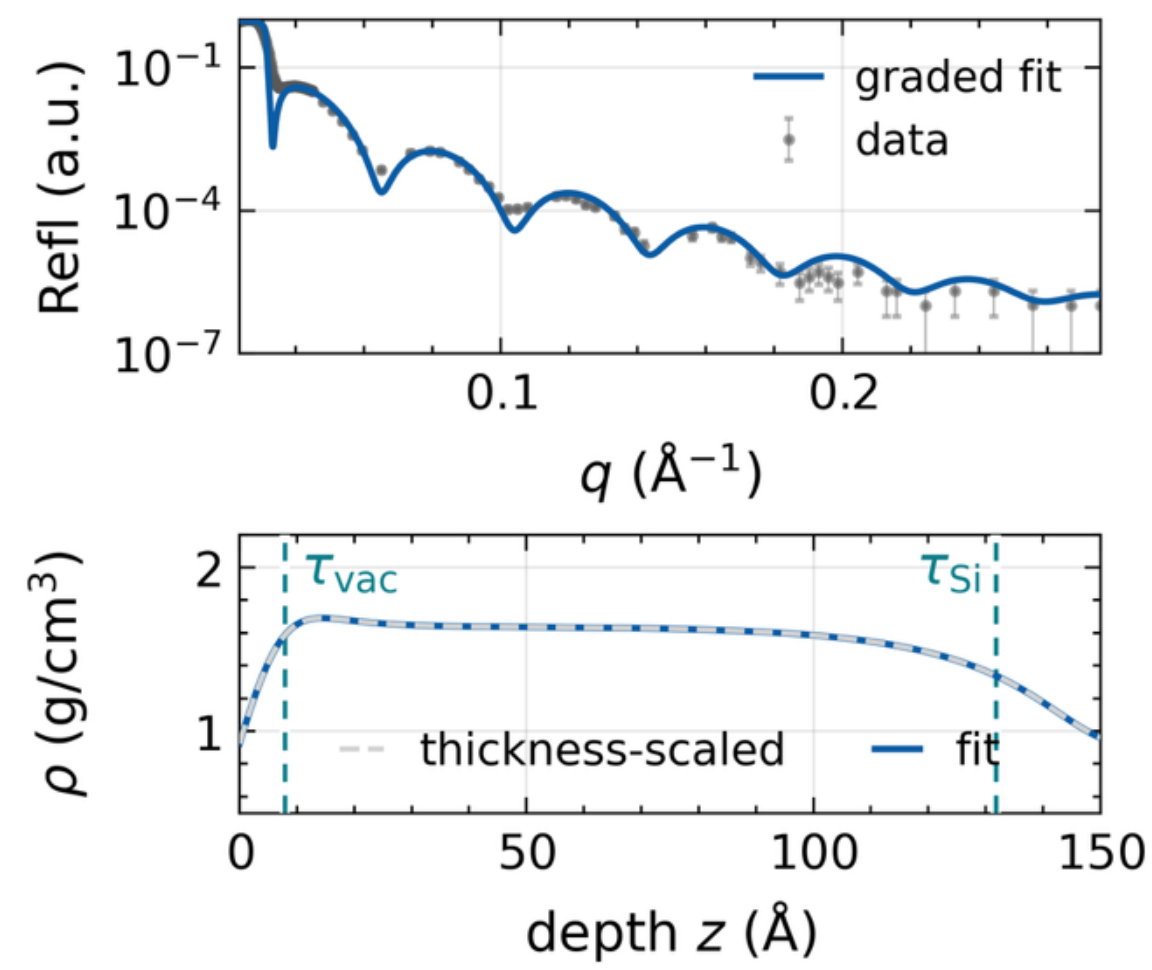

FIG. S10 Non-resonant and non-polarized XRR measured on a similarly prepared sample with thickness 15 nm measured by QCM and grown under otherwise identical conditions. Data was measured at Washington State University on a Xenocs Xeuss 3.0 with a focused Cu X-ray source. Reflectivity was fit using the starting parameters of the gradient model Table I while fitting only total film thickness. Once converged all other parameters where refined, $\rho_v$ , $\rho_b, \rho_{Si}, \tau_v, \tau_{Si}$, to less than $1\sigma$ from their initial position. A single non resonant energy is incapable of independently determining the orientation gradient, yet the density gradient is consistent with this second sample.

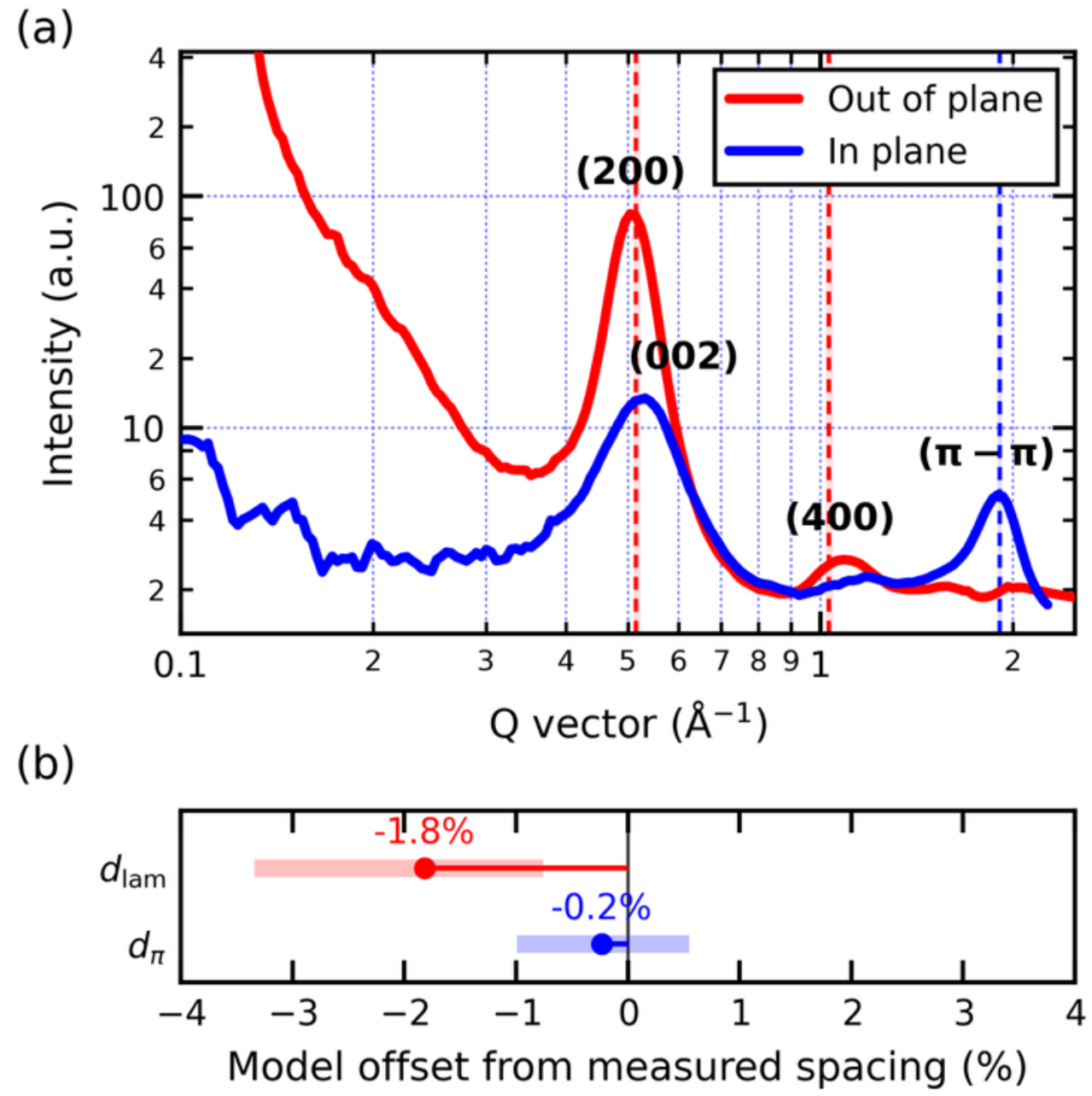

FIG. S11 Reconstructed packing (**Fig 4c**) compared with GIWAXS. **(a)** Out-f-plane (red) and in-plane (blue) sector cuts from **Fig S2** with lamellar orders and the $\pi$-stacking reflection indexed. Dashed lines mark the d-spacing predicted by the packing in **Fig. 4c**. **(b)** Comparison between predicted d-spacing and the measured d-spacing for both periods.

TABLE S III Unit cell parameters for the crystal phases used in MD simulations. Literature $\beta$-ZnPc unit cell parameters presented and rounded for comparison.[58] GIXRD constrained unit cell (EXP) compared with the (MD) relaxed cell, with uncertainties set by 10ps block averages over the final 50 ps of the simulation. Atomic coordinates for constructed $\alpha$ cells are provided as CIFs.

| | $\boldsymbol{a}$ (Å) | $\boldsymbol{b}$ (Å) | $\boldsymbol{c}$(Å) | $\boldsymbol{\beta}$(°) | $\boldsymbol{\rho(g/cm^3)}$ |
|---|---|---|---|---|---|
| $\boldsymbol{\beta}$ | 14.5 | 4.9 | 17.2 | 106.20 | 1.648 |
| $\boldsymbol{\alpha}$ (EXP) | 24.93 | 4.18 | 23.62 | 90.4 | 1.558 |
| $\boldsymbol{\alpha}$ (MD) | 24.647(3) | 3.917(2) | 23.985(10) | 90 | 1.657 |

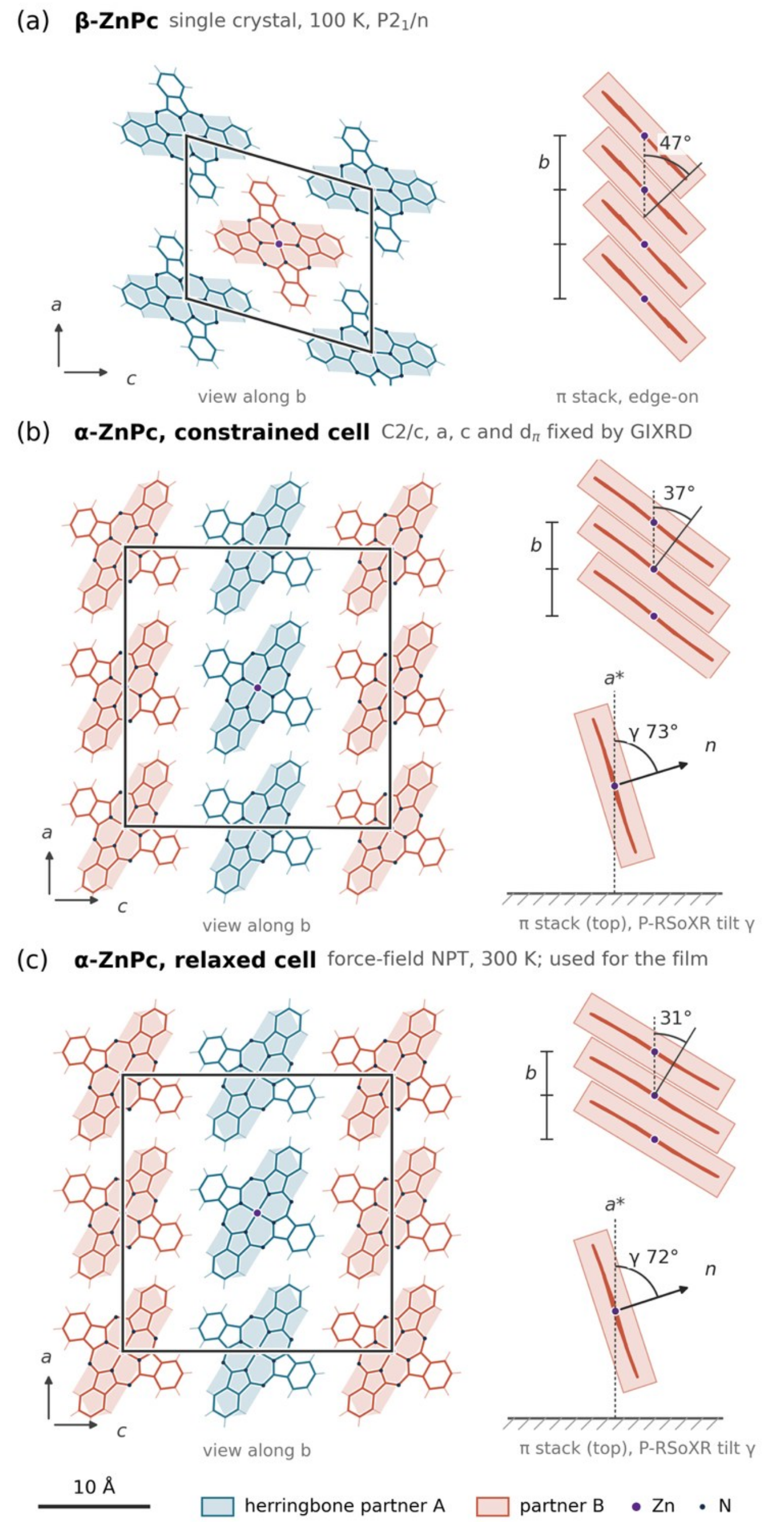

FIG. S12 Constructed unit cells for MD simulations. (a) $\beta$ phase ZnPc from the crystallography used for DFT simulation. (c) $\alpha$ phase ZnPc constructed from CuPc $\alpha$ cell, with spacings constrained from GIXRD. (d) Thermally relaxed $\alpha$-ZnPc cell through LAMMPS simulation. The $\alpha$-ZnPc cell is constructed on a repulsive wall and annealed at 300K.

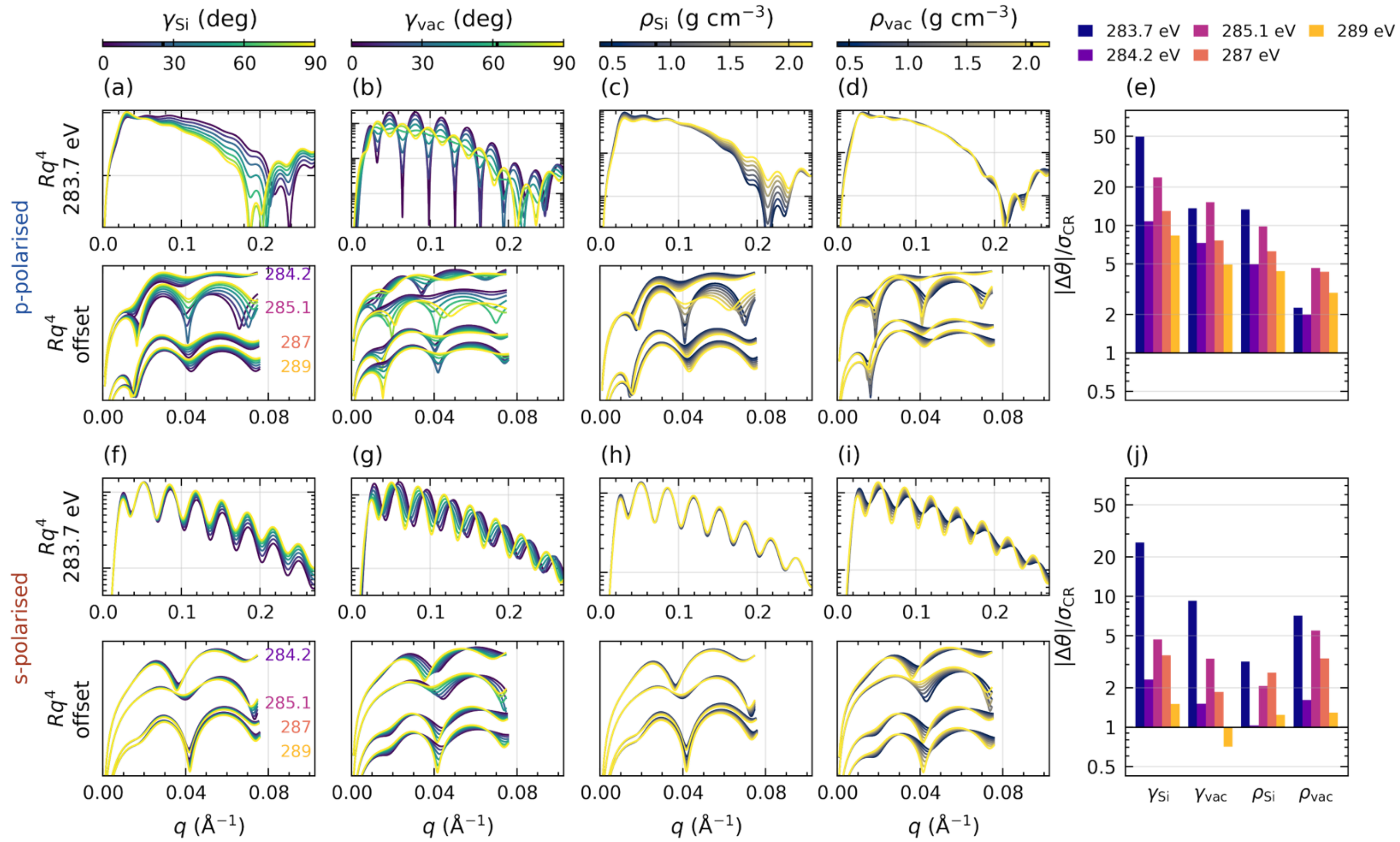

FIG. S13 Fisher sensitivity analysis for reflectivity of 2nd order orientation and density phase transitions for ZnPc. From the best fit model (**Fig 4a**), each pinned boundary is systematically varied along a feasible grid with characteristic thicknesses held constant. (**a-d**) p-polarized and (**f-i**) s-polarized simulated reflectivity ($Rq^4$). The upper axis of each reflectivity panel originates from the 283.7 eV diagnostic energy over the full measurement range, with the lower being the resonant series on the reduced q-range vertically offset for clarity. The reflectivity slope and fringe depth, that usually set by surface or interfacial roughness, are polarization and energy dependent. Moreover, the primary channel for sensitivity to the buried interface is the p-polarized reflectivity, especially for $q$ values near Brewster's angle causing 283.7 eV to dominantly dictate the buried microstructure. The Kiessig fringe phase is also set differently across energy and polarization by a combination by all four pinned interfaces with no single parameter fully setting the phase. (**e & j**) Detectability, $|\Delta\theta|/\sigma_{CR}$, for each dataset split by energy, parameter and polarization, where $\Delta\theta$ is the difference in each parameter from the bulk phase, and $\sigma_{CR}$ is the Cramér-Rao bound. The fisher information is evaluated at the measured $q$-points over every parameter of the joint fit along with the four boundary values, with the other parameters held constant. The measurement error is modified with a fractional error floor setting the $\chi_\nu^2 \to 1$ such that the best fit solution in the main text is treated as the ground truth. Bars raise from one standard error $|\Delta\theta|/\sigma_{CR} = 1$. The pinned silica state rises 37 and 9 standard errors from the bulk using a residual based sandwich covariance ($\sigma_{\gamma_{Si}} = 1.28°$, $\sigma_{\rho_{Si}} = 0.084$g/cm$^3$).